\documentstyle[aps,prl,eqsecnum,preprint,tighten]{revtex}
\begin{document}
\title
{Landau theory of quantum spin glasses of rotors and Ising spins}
\author{N. Read and Subir Sachdev}
\address{Department of Physics, P.O. Box 208120, Yale University,
New Haven, CT 06520-8120\\
and\\
Department of Applied Physics, P.O. Box 208284, Yale University,
New Haven, CT 06520-8284}
\author{J. Ye}
\address{Department of Physics, Harvard University, Cambridge MA 02138}
\date{December 6, 1994}
\maketitle
\begin{abstract}
We consider quantum rotors or Ising spins in a transverse field
on a $d$-dimensional lattice,
with random, frustrating, short-range, exchange interactions. The quantum
dynamics are associated with a finite moment of inertia for the rotors,
and with the transverse field for the Ising spins. For a suitable
distribution of
exchange constants, these models display spin glass and quantum paramagnet
phases
and a zero temperature quantum
transition between them. An earlier exact
solution for the critical properties of a model with infinite-range
interactions can be reproduced by
minimization of a Landau effective-action functional for the model in
finite $d$ with short-range
interactions. The functional is expressed in terms of a composite spin
field which is bilocal in time.
The mean-field phase diagram near the zero temperature critical point is
mapped out as a
function of temperature, strength of the quantum coupling, and applied
fields. The spin glass ground
state is found to be replica symmetric, with replica symmetry breaking
appearing only at finite
temperatures. Next we examine the consequences of fluctuations about
mean-field for the critical
properties. Above
$d=8$, and with certain restrictions on the values of the Landau couplings,
we find that the
transition is controlled by a Gaussian fixed point with mean-field critical
exponents. For
couplings not attracted by the Gaussian fixed point above $d=8$, and for
all physical couplings
below $d=8$, we find  runaway renormalization
group flows to strong coupling. General scaling relations that should be
valid even at the strong
coupling fixed point are proposed and compared with Monte Carlo simulations.
\end{abstract}
%\narrowtext

\tableofcontents

\section{Introduction}
\label{introduction}
Models of quantum rotors or the Ising model in a transverse field on a
$d$-dimensional lattice are among the simplest systems which exhibit a zero
temperature ($T$) quantum transition~\cite{kogut,chn,csy}.  This is
particularly
true in the absence of quenched disorder: then the critical properties map
directly onto those of the corresponding classical spin model in $d+1$
dimensions, with the quantum time behaving just like another spatial dimension.
However, in the presence of quenched disorder, the situation is more
complicated.
The disorder has no dynamic fluctuations, so the corresponding classical spin
models have randomness which is constant along the `time' direction;
unfortunately, there is only a rather limited existing body of knowledge on
such
classical systems that one can draw on.  Nevertheless, these models still
constitute an attractive setting for the study of the complicated interplay of
quenched disorder, interactions and quantum mechanics.
One can quite reasonably hope that the insight obtained from their analysis
might
prove profitable in other systems
which involve the same ingredients---these include the metal-insulator and
superfluid-insulator transitions and have been the focus of a large number of
experimental investigations.
Direct experimental studies of the models studied in this paper have been much
more limited, although a recent investigation of a system which can be
well described by the Ising model in a transverse field is
noteworthy~\cite{rosen}.
Three component quantum rotors with quenched disorder may also be a
reasonable starting
point for understanding some of the spin-fluctuation properties of the
doped cuprate compounds
in a regime with localized holes~\cite{qcrit}.

We consider the following Hamiltonian of quantum rotors on the
site $i$ of a regular, $d$-dimensional lattice:
\begin{equation}
{\cal H}_R = \frac{g}{2} \sum_i \hat{{\bf L}}_i^2 - \sum_{<ij>} J_{ij}
\hat{{\bf n}}_i \cdot
\hat{{\bf n}}_j
\label{hamrot}
\end{equation}
where the sum $<ij>$ is over nearest neighbors, although our results should
also
apply to models with more general short-range interactions. The $M$-component
vectors $\hat{{\bf n}}_i$, with $M \geq 2$, are of unit length, $\hat{{\bf
n}}_i^2 = 1$,
and represent the orientation of the rotors on the surface of a sphere in
$M$-dimensional rotor space. The operators $\hat{L}_{i\mu\nu}$ ($\mu < \nu$,
$\mu$, $\nu = 1 \ldots M$) are the $M(M-1)/2$ components of the angular
momentum
$\hat{{\bf L}}_i$ of the rotor: the first term in ${\cal H}_R$ is the
kinetic energy of the
rotor with $1/g$ the moment of inertia. The different components of $\hat{n}_i$
constitute a complete set of commuting observables and the state of the system
can be described by a wavefunction $\Psi ({n}_i)$.
The action of $\hat{{\bf L}}_i$ on $\Psi$ is given by the usual
differential form of the
angular momentum
\begin{equation}
L_{i\mu\nu} = -i \left( n_{i\mu} \frac{\partial}{\partial n_{i\nu} }
- n_{i \nu} \frac{\partial}{\partial n_{i \mu} } \right).
\end{equation}
The commutation relations among the $\hat{{\bf L}}_i$ and $\hat{{\bf n}}_i$
can now be easily
deduced. We emphasize the difference of the rotors from Heisenberg-Dirac
quantum spins:
the components of the latter at the same site do not commute, whereas the
components of the
$\hat{{\bf n}}_i$ do.

Let us also introduce the
Hamiltonian of the Ising model in a transverse field, ${\cal H}_I$:
\begin{equation}
{\cal H}_I = -g \sum_i \sigma_{i}^{x} - \sum_{<ij>} J_{ij} \sigma_{i}^z
\sigma_j^z .
\label{hamising}
\end{equation}
Here $\sigma^x , \sigma^z$ are the $x,z$ components of the three Pauli spin
operators, with the Pauli operators on different sites commuting with each
other.
Each site, therefore, has an Ising degree of freedom, represented by the
eigenvalues of the $\sigma_i^z$. The $\sigma_i^x$ is the kinetic energy term
and induces on-site flips of the Ising spins.

The analogy between ${\cal H}_R$ and ${\cal H}_I$ should be quite clear.
${\cal H}_R$ has a global
${\rm O}(M)$ symmetry, while ${\cal H}_I$ has a global, $Z_2$, spin-flip
symmetry associated
with the unitary transformation $U = \prod_i \sigma_i^x $. In both models, the
interaction terms proportional to the $J_{ij}$ would, if $g$ were zero,
prefer a ground state
in which each rotor or Ising spin has a definite orientation which
minimizes the exchange
energy, and quantum fluctuations are absent: any such choice will break the
global ${\rm O}(M)$ or $Z_2$ symmetry in a given sample (although the
spin-glass
phase preserves a statistical spin symmetry after averaging, discussed in
Sec~\ref{spinglassphase}). A small $g$ will not necessarily destroy this
phase, in sufficiently
high spatial dimension. In the opposite
limit $J_{ij} =0$, both ${\cal H}_R$ and ${\cal H}_I$ possess non-degenerate
ground states which preserve the global symmetry. For the rotors, each site is
in the spherically symmetric `$s$-wave' state (using the language of $M=3$).
Similarly, each Ising spin is in the eigenstate of $\sigma^x$ with
eigenvalue $+1$:
$= (|\uparrow\rangle + |\downarrow\rangle)/\sqrt{2}$ which is $Z_2$ invariant.
In both cases, the $J_{ij}=0$ ground state is separated from the
first excited state by a gap of order $g$. It is therefore reasonable to
consider the
$J_{ij}$'s as perturbations in this limit (though the random nature of the
$J_{ij}$ causes
some problems which we will discuss later) and to expect that this
``quantum-disordered''
phase persists at finite values of $J_{ij}$. Thus we expect both ordered and
disordered phases to exist at zero temperature.

The above discussion should make it clear that it is natural to consider the
transverse-field Ising model as simply the $M=1$ case of the quantum rotors.
As in the classical and non-random quantum cases, the main difference between
$M=1$ and
$M > 1$ is that the latter has a continuous symmetry and possesses gapless
spin-wave excitations in magnetically ordered phases. For ease of the
subsequent
discussion, it is convenient to introduce a notation which allows simultaneous
discussion of the rotor and Ising models. We therefore introduce an
$M$-component
spin $S_{i\mu}$, with $M\geq 1$, such that
\begin{equation} S_{i\mu} = \left\{ \begin{array}{c}
\mbox{$\hat{n}_{i\mu}$ for $M \geq 2$} \\
\mbox{$\sigma^z_i$ for $M=1$}.
\end{array}
\right.
\end{equation}
We will use the general $S_{i\mu}$ notation in most of the remainder of the
paper.

We now want to describe the possible phases of ${\cal H}_R$ or ${\cal H}_I$
at zero
temperature. The
$J_{ij}$ are assumed to be statistically independent between different links
and
to possess the following first and second moments:
\begin{equation}
\left[J_{ij}\right] = J_0~~~~;~~~~\left[(J_{ij} - J_0 )^2\right] = J^2 .
\end{equation}
Here, and henceforth, the square brackets will represent an average over the
quenched disorder. Averages over quantum or thermal fluctuations will be
represented by angular brackets. On bipartite lattices, the properties of the
system are invariant under the global sign change $J_{ij} \rightarrow -
J_{ij}$:
this is because the sign of the exchange energy can be reversed by a global
spin-flip $S_{i\mu} \rightarrow - S_{i\mu}$ on one sublattice. On such lattices
we can therefore assume without loss of generality that $J_0 > 0$, and we will
so assume.  Note that this property is not shared by models where
$S_{i\mu}$ is replaced by quantum Heisenberg spins whose components do not
commute.

It is also possible to modify ${\cal H}_R$ and ${\cal H}_I$ by allowing $g$
to have random
fluctuations about its mean value. This does not affect the analysis of the
phases of the model. It does represent an important type of randomness that
will
be considered in our discussion of the Landau theory.

We have now introduced three energy scales, $g$, $J_0$ and $J$,
and the nature of the ground state becomes especially clear when one of the
three
scales is much larger than the other two. The three phases so obtained are
the ferromagnet ($J_0 \gg g, J$), the spin-glass ($J \gg J_0, g$),
and the quantum paramagnet ($g \gg J_0, J$). We review the structure of the
three phases in turn (analogous to finite temperature phases in classical
models):
\paragraph{Ferromagnet:}
Each site acquires a static moment, and further the average of the moments
on the
different sites is non-zero:
\begin{equation}
\left\langle S_{i\mu} \right\rangle \neq 0 ~~~~;~~~~\left[\left\langle
S_{i\mu} \right\rangle\right] = M_{0\mu} \neq 0.
\end{equation}
The ferromagnetic order parameter is $M_{0\mu}$; note however that the
first, quantum expectation value will have fluctuations from site to site about
$M_{0\mu}$. The reader may be aware of mappings between the low-energy
properties
of clean quantum Heisenberg spin models and quantum rotors~\cite{haldane}:
we note
here that it is the {\it anti\/}ferromagnetically-ordered N\'{e}el phase of
Heisenberg spin that maps onto the ferromagnetic phase of quantum rotors.

\paragraph{Spin-glass:} Each site now has a random static
moment~\cite{youngbinder,book}, and
the average moment is therefore zero:
\begin{equation}
\left\langle S_{i\mu} \right\rangle \neq 0 ~~~~;~~~~\left[\left\langle
S_{i\mu} \right\rangle\right] = 0.
\end{equation}
The Edwards-Anderson order parameter, $q_{EA}$, is given by
\begin{equation}
q_{EA} = \frac{1}{M} \lim_{\tau\rightarrow \infty}
\left[\left\langle S_{i\mu} ( 0 ) S_{i\mu} ( \tau ) \right\rangle \right].
\label{defqea1}
\end{equation}
(We are using here the Einstein summation convention on the ${\rm O} (M)$
vector
index $\mu$; this convention will be implicitly assumed throughout the paper.)
One would expect that $q_{EA}$ is also equal to
\begin{equation}
q = \frac{1}{M}
\left[\left\langle S_{i\mu} \right\rangle^2 \right].
\label{defqea2}
\end{equation}
In classical spin glasses, $q \neq q_{EA}$ when replica symmetry breaking takes
place~\cite{youngbinder,book}, because the $\langle \cdot \rangle$ average
must be carefully
defined. In spite of the quantum fluctuations that are present, the
appearance of the
ferromagnetic and spin glass phases is quite similar to that of their classical
analogues at $T=0$ or $T \neq 0$.
\paragraph{Quantum paramagnet:}
This phase has no static moment:
\begin{equation}
\left\langle S_{i\mu} \right\rangle = 0.
\end{equation}
It differs in one important respect from the strong-coupling, $J_{ij} = 0$,
picture discussed earlier. The gap in the excitation spectrum is filled in
by contributions from rare regions in which the local values of $J_{ij}$ are
such as to place the system in one of the magnetically ordered phases. A
brief discussion of such `Griffiths effects' is presented in
Appendix~\ref{griffiths}.

There has been some earlier work on the ferromagnetic and
paramagnetic phases and the transition between them. In one dimension, many
exact
properties have been determined for the transverse field Ising
model~\cite{wu,shankar,daniel}.  In $d=1$, the $M=1$ spin-glass phase and its
transitions are closely related to those of the ferromagnet, as the former
can be
related to the latter by a gauge transformation which makes all the $J_{ij}$
positive. (The magnetically-ordered phases do not exist for $M \geq 2$ in
$d=1$).
In higher dimensions, there has been a discussion of the field-theoretic
properties of the ferromagnetic-paramagnetic
transition~\cite{doro,cardy,lawrie}.
Boyanovsky and Cardy~\cite{cardy} identified $d=4$ as the upper-critical
dimension
and discussed the scaling properties of the correlation functions. A study
of the
critical properties below $d=4$ in an expansion in
$\varepsilon=4-d$ yields a flow to strong-coupling with no stable, physical
fixed
point (the only stable fixed-point has unphysical properties like a negative
variance of some observables). However, Boyanovsky and Cardy~\cite{cardy}
showed
that a stable, physical, fixed point could be obtained in a theory with
$\epsilon_{\tau}$ time dimensions, in a double expansion in
$\varepsilon$ and $\epsilon_{\tau}$; it is not known whether this fixed point
continues to be pertinent all the way to $\epsilon_{\tau} = 1$.

We will not make any further references to the ferromagnetic phase. It is
convenient, therefore, to specialize henceforth to the case of $J_{ij}$
distributions which are symmetric about $J_{ij} = 0$ and in particular have
\begin{equation}
J_0 = 0.
\end{equation}
In this case, it can be shown~\cite{book} that the disordered-averaged
correlation functions are in fact invariant under a $Z_2$ gauge transformation
$S_{i \mu} \rightarrow \eta_i S_{i \mu}$, with $\eta_i = \pm 1$ and
site-dependent.  A zero temperature spin glass to quantum paramagnet phase
transition will occur
as the ratio $g/J$ is increased and is the subject of most of the following.

The spin-glass phase and its transitions were studied in
Refs~\cite{BrayMoore,Gold},
with considerable additional interest in the last
year~\cite{miller,letter,guo,rieger,rieger2,opperman,thill}. Huse and
Miller~\cite{miller} studied
the transverse-field Ising model with infinite-range interactions and
determined exact critical
properties of the transition separating the spin-glass and paramagnetic
phases. We have
studied~\cite{letter} the  quantum-rotor model with infinite-range
interactions and obtained
essentially identical results. The independence of the critical properties
on the value of $M$ was
also understood~\cite{letter}. We now highlight features of these results
which will be
important for our discussion.

\subsection{Review of results for the infinite range model}
\label{infiniterange}
In the infinite-range version of ${\cal H}_R$ (\ref{hamrot}), or ${\cal
H}_I$ (\ref{hamising}),
the interactions $J_{ij}$ are taken to be independent random variables with
distribution
\begin{equation}
P(J_{ij} ) \propto \exp\left(- N J_{ij}^2 / 2 J^2  \right)
\end{equation}
for all pairs $i,j$ of distinct sites; $N$ is the number of sites. Note that
the variance of $J_{ij} \sim J^2 /N$ in the infinite range model, but is
$\sim J^2$
in the finite range model. As $N\rightarrow \infty$ mean-field theory becomes
exact, which means that one only need solve a problem of a single site with a
self-interaction between different times $\tau_1, \tau_2$ of $J^2 D (\tau_1
- \tau_2 )$, where $D$ has to be determined self-consistently as
\begin{equation}
D( \tau_1 - \tau_2 ) = \frac{1}{M} \left\langle S_{\mu} ( \tau_1 ) S_{\mu}
( \tau_2 )
\right\rangle_{D}
\end{equation}
with the average being calculated with the same self-interaction $D$.
This was done by self-consistency arguments on the spectral density by
Miller and
Huse, and by taking the limit $M \rightarrow \infty$ and then expanding in
$1/M$
using similar arguments, by the present authors. $D$ represents the disorder
average
\begin{equation}
D( \tau_1 - \tau_2 ) = \frac{1}{M} \left[\left\langle S_{i\mu} ( \tau_1 )
S_{i\mu} ( \tau_2 )
\right\rangle\right]
\end{equation}
in the original random problem. It was found that a spin-glass to paramagnet
transition occurred at $g=g_c \sim J$. At $g > g_c$, $D$ decays
exponentially with
$\tau$ as $\tau \rightarrow \infty$, indicating a gap $\Delta$ in the
corresponding spectral density; at $g=g_c$, $D$ decays as $1/\tau^2$, and in
the
ordered phase, $D \rightarrow$ constant $= q_{EA}$. The Fourier transform $D (
\omega )$ of $D ( \tau )$ has the form
\begin{equation}
D ( \omega ) \sim \mbox{constant} - \sqrt{ \omega^2 + \Delta^2 }
\label{valDinf}
\end{equation}
for $g \geq g_c$, which is responsible for the $1/\tau^2$ behavior at $g_c$
and it
turns out that the correlation time $\xi_{\tau} \sim \Delta^{-1}$ diverges as
\begin{equation}
\xi_{\tau} \sim \left( (g - g_c )^{-1} \log (g - g_c ) \right)^{1/2}.
\end{equation}
Thus we can define an exponent $z \nu$, anticipating anisotropic scaling in
space
and time in the short-range model, which takes the value $z \nu = 1/2$ in the
infinite range model. The log correction originates from effects of the length
constraint. Its significance will be fully explained in our later analysis.
Since $D ( \tau \rightarrow \infty )$ is the Edwards-Anderson order
parameter, we
may also define $q_{EA} = (g_c - g)^{\beta}$ and it is found that $\beta = 1$.
At $g=g_c$ one expects $D (\tau ) \sim \tau^{-\beta /z\nu}$~\cite{qcrit}
which is satisfied
with the values already obtained.

It is perhaps surprising that the critical properties of the self-consistent
single-site quantum problem can be obtained exactly for all $M$, and it
remains to
explain some features of the results, such as the logarithmic violations of
scaling. One of our goals in this paper is to provide a conceptually simpler
derivation of the results based on a Landau action functional and to explain
the
logarithms as due to the decay of a marginally irrelevant variable which we
will
identify. We also wish to calculate critical properties in short-range,
finite-dimensional models, and to this end we will study mean-field theory and
Gaussian fluctuations around it in the Landau theory framework. We will
show that
in this approximation, $z=2$ and $\nu = 1/4$. We will also define correlation
functions in the next subsection and obtain their scaling properties in the
same
approximation later, obtaining $\eta = 0$ by definition in mean-field theory.
For sufficiently large $d$ ($d > 8$) and for a certain range of values of
Landau
couplings, we will show that these mean field theory results are valid and
have the same critical
properties as in the infinite-range model (where comparisons are possible,
{\em i.e.\/} not for
correlations at large spatial separation and not for the non-linear
susceptibility which
has a singular infinite-range limit~\cite{private_huse}). However outside this
range, and for all couplings at $d < 8$, we find a renormalization group
flow to a
region where perturbation theory breaks down.

\subsection{Order parameter and observables}
\label{observables}
In classical finite temperature transitions the equilibrium critical properties
can be studied in terms of static fluctuations of some ``order-parameter''
field.
For classical spin glasses in the replica formalism, this is a matrix $q^{ab}$,
$a,b =1 \ldots n$ are replica indices and $n \rightarrow 0$.
The {\em off\/}-diagonal components of $q_{ab}$ can be related to the
Edwards-Anderson order parameter in a somewhat subtle way we won't go into
here~\cite{youngbinder,book}. In quantum ($T=0$) phase transitions, time
dependent
fluctuations of the order parameter must be considered (in ``imaginary''
Matsubara time $\tau $) and in the spin glass case it is found that the
standard
decoupling, analogous to the classical case introducing $q^{ab}$ leads now to a
matrix function of two times~\cite{BrayMoore} (see also
Appendix~\ref{derivation})
which we can consider to be
\begin{equation}
Q_{\mu\nu}^{ab} ( x , \tau_1 , \tau_2 ) =  \sum_{i \in {\cal N}(x)}
S_{i\mu}^a ( \tau_1 )
S_{i \nu}^b (\tau_2
)
\label{composite}
\end{equation}
where ${\cal N}(x)$ is a coarse-graining region in the neighborhood of $x$.
The definition (\ref{defqea1})
of the Edwards-Anderson order parameter now implies
\begin{eqnarray}
D ( \tau_1 - \tau_2 ) & \equiv &
\lim_{n \rightarrow 0}
\frac{1}{Mn} \sum_{a} \left\langle\left\langle Q^{aa}_{\mu\mu} ( x, \tau_1 ,
\tau_2) \right\rangle\right\rangle \label{defD} \\
q_{EA} &=& \lim_{\tau \rightarrow \infty} D( \tau )
\label{defqea}
\end{eqnarray}
relating $q_{EA}$ to the replica diagonal components of $Q$. (A reminder:
the Einstein summation
convention on the vector index $\mu$ is being used in (\ref{defD}).
However, no such convention is used for the $a,b$ replica indices). It is
important to note here that the definition (\ref{defqea2})  also relates
$q_{EA}$
to the replica off-diagonal components of $Q$~\cite{book}, whose
expectation value will be
time independent.
We have introduced above double angular brackets to represent averages
taken with
the translationally invariant replica action (recall that single angular
brackets
represent thermal/quantum averages for a fixed realization of randomness,
and square brackets represent averages over randomness).

We saw in the infinite range model that the behavior of $D$ at long times
in mean
field theory changed significantly at the transition and it is clear that the
theory of fluctuations must include the whole matrix function $Q$, for {\em
both\/} its diagonal and off-diagonal components. Strictly speaking the replica
diagonal $Q$ at finite $\tau_1 - \tau_2$ is not an order parameter because its
expectation is non-zero on both sides of the transition; nonetheless it
does play
the role of the ``order parameter field'' in the Landau theory.

In the finite dimensional model, where $J_{ij}$ couples nearest neighbors only,
$D(\tau)$ is expected to acquire a long time limit only at the transition
to, and
in, the spin glass phase. However its behavior in the paramagnetic phase may
differ from that in mean field theory (the infinite range model) for the
following reason. Since the $J_{ij}$'s are random, there is some chance of any
given region having all $J_{ij}$'s large in magnitude and thus resembling a
patch
of the ordered phase. Such an event would be statistically rare, but can
contribute significantly to the long time behavior of $D$ (which is the average
on-site correlation function) because the finite region will have slow
overturns
of its instantaneous moment. The effects of these ``Griffiths singularities''
on
$D$ has been discussed recently by Thill and
Huse~\cite{thill} for
$M=1$; the extension of their argument to $M > 1$ is presented in
Appendix~\ref{griffiths};  for $M > 1$, we find only weak essential
singularities at $\omega=0$ in the spectral function for $D$.
An important question is whether similar fluctuations  affect
the critical properties, an effect presumably not included in the treatment
that
will be given later.

It is also helpful to introduce here a number of correlation functions of
the order
parameter whose scaling properties will be described in the paper.
A quantity intimately related to the spin glass long range order is the quantum
mechanical disconnected correlation function
\begin{equation}
G ( i-j, \tau_1 - \tau_2 , \tau_3 - \tau_4 ) \equiv
\left[ \left\langle S_{i\mu} (\tau_1 ) S_{j\mu} (\tau_2 ) \right\rangle
\left\langle S_{i\nu} (\tau_3 ) S_{j\nu} (\tau_4 ) \right\rangle \right].
\end{equation}
Note that $\left[\left\langle S_{i\mu} ( \tau_1 ) S_{j\mu} ( \tau_2 )
\right\rangle
\right] = 0$ for $i\neq j$ because of the $Z_2$ gauge symmetry, and no
subtraction of
products of disorder averages is necessary, as
a subtraction analogous to that in (\ref{defgd1}) below will vanish for
this case.
After coarse-graining both $i$ and $j$ over their respective averaging regions
in the neighborhoods of $x$ and $y$, we
obtain the correlator of the order parameter $Q$
\begin{equation}
G (x-y, \tau_1 - \tau_2 , \tau_3 - \tau_4 ) =  \lim_{n \rightarrow 0}
\frac{1}{n(n-1)} \sum_{a\neq b} \left\langle\left\langle Q^{ab}_{\mu\nu}
(x, \tau_1 , \tau_3 ) Q^{ab}_{\mu\nu} ( y, \tau_2 , \tau_4 )
\right\rangle\right\rangle.
\label{defgd2}
\end{equation}
$G$ will be found later to behave as the propagator for fluctuations of
the $Q$ order parameter field about the mean-field theory, and is directly
analogous to a
corresponding object in the classical spin glass.

A second correlator arises upon
considering fluctuations (due to the randomness) of the on-site spin
correlation function
$\left\langle S_{i\mu} ( \tau_1 ) S_{i \mu} ( \tau_2 ) \right\rangle$. The
second cumulant
of these fluctuations can be obtained from a quantum mechanically disconnected
correlation function
\begin{eqnarray}
G^{d} ( i-j, \tau_1 - \tau_2 , \tau_3 - \tau_4 ) \equiv &&
\left[ \left\langle S_{i\mu} (\tau_1 ) S_{i\mu} (\tau_2 ) \right\rangle
\left\langle S_{j\nu} (\tau_3 ) S_{j\nu} (\tau_4 ) \right\rangle \right]
\nonumber \\
&&~~~~~~~~~~- \left[ \left\langle S_{i\mu} (\tau_1 ) S_{i\mu} (\tau_2 )
\right\rangle
\right]\left[
\left\langle S_{j\nu} (\tau_3 ) S_{j\nu} (\tau_4 ) \right\rangle \right]
\label{defgd1}
\end{eqnarray}
After coarse graining this becomes another two-point correlation function
of the order
parameter, $Q$,
\begin{eqnarray}
G^{d} (x-y, \tau_1 - \tau_2 , \tau_3 - \tau_4 ) = && \lim_{n \rightarrow
0}
\frac{1}{n(n-1)} \sum_{a\neq b} \left\langle\left\langle Q^{aa}_{\mu\mu}
(x, \tau_1 , \tau_2 ) Q^{bb}_{\nu\nu} ( y, \tau_3 , \tau_4 )
\right\rangle\right\rangle \nonumber\\
&&~~~~~~~~~~~~~~~~~~~~~~~~ - D(\tau_1 - \tau_2 ) D( \tau_3 - \tau_4 )
\end{eqnarray}
obtained as before by averaging over $i$ and $j$ in neighborhoods of $x$
and $y$.
The analog of $G^d$ in a classical spin glass is trivial, since $S_{i\mu}^2
= 1$.
Higher moments of the on-site correlation function can also be constructed,
and the entire set is expected to have rather non-trivial scaling
properties near
the quantum phase transition~\cite{ludwig}.

Finally, to exhaust the set of different two-point correlators of the
$Q$ field, we consider the connected correlation function $G^c$
\begin{eqnarray}
G^c_{\mu\nu\rho\sigma}
(i - j , \tau_1 - \tau_4 , \tau_2 - \tau_4 &&, \tau_3 -
\tau_4) =  \left[\left\langle S_{i\mu} ( \tau_1 ) S_{i\nu} (\tau_2 )
S_{j \rho} ( \tau_3 ) S_{j \sigma} ( \tau_4 ) \right\rangle\right] \nonumber \\
&&~~- \frac{1}{M^2} \delta_{\mu\nu}
\delta_{\rho\sigma} \left[\left\langle S_{i\alpha} ( \tau_1 ) S_{i\alpha}
(\tau_2
)\right\rangle\left\langle  S_{j \beta} ( \tau_3 ) S_{j \beta} ( \tau_4 )
\right\rangle\right]
\nonumber \\
&&~~- \frac{1}{M^2}
\delta_{\mu\rho}\delta_{\nu\sigma} \left[\left\langle S_{i\alpha} ( \tau_1
) S_{j\alpha} (\tau_3
) \right\rangle\left\langle S_{i \beta} ( \tau_2 ) S_{j \beta} ( \tau_4 )
\right\rangle\right]
 \nonumber \\
&&~~ - \frac{1}{M^2} \delta_{\mu\sigma}
\delta_{\nu\rho} \left[\left\langle S_{i\alpha} ( \tau_1 ) S_{j\alpha} (\tau_4
)
\right\rangle\left\langle S_{i \beta} ( \tau_2 ) S_{j \beta} ( \tau_3 )
\right\rangle\right],
\label{defgc}
\end{eqnarray}
where we used the symmetry properties.
After coarse-graining, we obtain
\begin{eqnarray}
G^c_{\mu\nu\rho\sigma}
(x - y ,&& \tau_1 - \tau_4 , \tau_2 - \tau_4 , \tau_3 -
\tau_4) = \lim_{n \rightarrow 0}
\frac{1}{n} \sum_{a} \left\langle\left\langle Q^{aa}_{\mu\nu}
(x, \tau_1 , \tau_2 ) Q^{aa}_{\rho\sigma} ( y, \tau_3 , \tau_4 )
\right\rangle\right\rangle \nonumber \\
&&~~- \frac{1}{M^2} \delta_{\mu\nu}
\delta_{\rho\sigma} \left\{
G^{d} (x - y , \tau_1 - \tau_2 , \tau_3 - \tau_4 ) + D(\tau_1 -
\tau_2) D (\tau_3 - \tau_4 ) \right\} \nonumber \\
&&~~- \frac{1}{M^2}
\delta_{\mu\rho}\delta_{\nu\sigma} G (x - y, \tau_1 -
\tau_3 , \tau_2 - \tau_4 ) \nonumber \\
&&~~ - \frac{1}{M^2} \delta_{\mu\sigma}
\delta_{\nu\rho} G ( x - y ,
\tau_1 - \tau_4 , \tau_2 - \tau_3 ).
\end{eqnarray}
Any scaling theory of the transition should obtain the scaling dimensions
and functions of
$D$, $G$, $G^{d}$ and $G^{c}$, and the possibilities appear rather
varied.

Two important susceptibilities can also be related to the correlators
considered above. The Edwards-Anderson spin-glass susceptibility
$\chi_{sg}$ is given
in the paramagnetic phase by
\begin{equation}
\chi_{sg} = \sum_{j} \left[ \chi_{ij}^2 \right]
\end{equation}
where
\begin{equation}
\chi_{ij} = \int d \tau \left\langle S_{i\mu} ( 0) S_{j\mu} ( \tau )
\right\rangle.
\end{equation}
and is analogous to the corresponding object used in the classical
theories. After coarse graining we have the expression in terms of $G$
\begin{equation}
\chi_{sg} = \int d^d x d\tau_1 d \tau_2 G (x,
\tau_1, \tau_2 ).
\end{equation}
The second susceptibility associated with the spin glass order is the
non-linear
susceptibility, $\chi_{nl}$, which is given by
\begin{equation}
\chi_{nl} =  \int d^d x d \tau_1 d \tau_2 d \tau_3
G^c_{1111} (x, \tau_1 , \tau_2 , \tau_3 ).
\label{defchinl}
\end{equation}

The outline of the remainder of the paper is as follows. In
Section~\ref{landausec} we begin by
setting up the Landau action functional for the finite dimensional models
near the zero
temperature phase transition, on which most of our results will be based. In
Section~\ref{meanfieldtheory}, this functional is minimized to yield mean
field theory, including
for the first time external fields; correlation functions are discussed at
the Gaussian level.
In Section~\ref{renormgroup}, a renormalization group method is used to
examine the stability of
the mean-field results to fluctuations, and to search for non-trivial exponents
below the upper
critical dimension $d_u = 8$ (no accessible perturbative fixed points are
found in $d<8$, however).
Section~\ref{scalinghypoth} discusses general scaling theory and compares
with results from
numerical simulations. Sketches of a derivation of the Landau theory and of
Griffiths
singularities appear in the appendices.

\section{Landau theory}
\label{landautheory}
\subsection{Landau action functional}
\label{landausec}
In order to write down the Landau theory for our model it will be useful to
first review
what is required in general of such a theory. The starting point of a
Landau theory is a
Landau (or Landau-Ginzburg-Wilson) functional which as a first step
describes the free
energy of the system near its critical point as a functional of its order
parameter, for
example the magnetization $\phi$ of an Ising-like ferromagnet near its
classical finite
temperature transition. In principle, the functional arises by considering
the free energy
in the presence of a field, say $h(x)$, that is thermodynamically conjugate
to $\phi$
(so that $- \int h \phi d^d x$ appears in the Hamiltonian), finding the
expectation of
$\phi$ for each $h(x)$, and then writing the free energy as a functional of
$\phi$ through
a Legendre transformation. In the present example this would be assumed to
take the form
\begin{equation}
{\cal F} = \int d^d x \left[ \frac{1}{2} ( \nabla \phi )^2 + \frac{1}{2} r
\phi^2
+ \frac{1}{4} \lambda \phi^4 + \ldots \right]
\end{equation}
where the dots denote terms with higher powers of $\phi$ and/or more
derivatives. Mean field
(or Landau) critical behavior can be obtained by taking $\phi (x) = \phi$
independent of
$x$ and minimizing with respect to $\phi$. Note that $r \sim T - T_c$
appears linearly.
Then for $r > 0$, $\phi = 0$ and for $r < 0$ , $\phi = \pm
(|r|/\lambda)^{1/2}$, and the
usual critical behavior of Landau theory can be obtained.

The key assumptions of Landau's approach are that, since $\phi$ is zero on
the disordered
(paramagnetic) side, then we can consider $\phi$ small and expand in
powers, and the
lowest powers will dominate near the critical point (odd powers are dropped
by symmetry
in this particular example). Similarly dependence of the coefficients on
$T-T_c$ is
analytic and all except the leading one $r = T - T_c$ can be dropped. In
considering
position dependent fluctuations, ${\cal F}$ can be used as the action in a
functional
integral over $\phi$, and the interactions of $\phi$ at different positions are
represented by the expansion in powers of gradients, corresponding to the
short range
interactions in the original physical problem.

Turning to the quantum spin glasses discussed in
Section~\ref{introduction}, we notice as mentioned
there that the ``order parameter'' $\langle\langle Q_{\mu\nu} (x, \tau ,
\tau^{\prime}
)\rangle\rangle$ is found in mean field theory to be a {\em function\/} $
\delta_{\mu \nu}
D(\tau -
\tau^{\prime} ) $, independent of $x$, which is non-zero for all finite $\tau -
\tau^{\prime}$ even in the paramagnetic phase. We therefore cannot expand
in powers of
$Q$ as $Q$ is typically not small. If instead we attempt to expand in
powers of $Q- D$,
it turns out that the action obtained contains very nonanalytic frequency
dependence near $g=g_c$
and so is unsatisfactory
according to the criteria above.

The solution is found by examining the Fourier transform $D(\omega )$ of
$D(\tau )$, Eqn
(\ref{defD}). $D(\tau)$ is a positive function that decreases monotonically
with $|\tau |$,
and its Fourier transform has similar properties in $\omega$. The constant
in Eqn
(\ref{valDinf}) is thus of order 1, and in fact can contain dependence on
$g-g_c$ and on
$\omega$, but these are analytic and negligible compared with the more singular
dependence contained in $\sqrt{\omega^2 + \Delta^2}$. The latter part is
nonanalytic in
both $\omega$ and $g - g_c$ as $g \rightarrow g_c$ and $\omega \rightarrow
0$, but it
{\em is\/} small in this region, and it is this part in which we wish to
expand the free
action. That is, we expand in $Q ( x, \omega_1 , \omega_2 ) - \mbox{constant}
\delta ( \omega_1 + \omega_2 )$ where the constant is chosen so that
$\langle\langle Q
\rangle\rangle - \mbox{constant}$ is zero (only) at $\omega_1 = \omega_2
=0$ and $g=g_c$. The
nonanalytic behavior of $\langle\langle Q \rangle\rangle - \mbox{constant}$
away from this point
should emerge from minimizing the action that should be analytic in form in
both frequency (or
$\tau$ derivatives) and $g-g_c$. The $\log(g-g_c)$ corrections to power-law
scaling
should also emerge in this way.

In Appendix~\ref{derivation} we sketch a direct derivation of the action
functional in a
microscopic model. Here we will proceed by the alternative procedure of
writing down all
terms allowed by the above considerations, together with simple symmetry
requirements.
We use the same notation $Q(x, \tau_1 , \tau_2)$ for the new, shifted
field. As it
differs only by a $\delta(\tau_1  - \tau_2 )$ from the original field,
which has little
effect on the scaling of critical correlations, we can usually ignore the
change in
definition.

Explicitly, the terms allowed in the action must satisfy:
\begin{enumerate}
\item The action is an integral over space of a local operator which can be
expanded in
gradients of powers of $Q$ evaluated at the same position $x$.
\item $Q$ is bilocal ({\em i.e.}\ is a matrix) in time, and each time is
associated with
one of the two replica indices and with one of the two ${\rm O}(M)$ vector
indices (see
definition Eqn (\ref{composite})). All these ``indices'' can appear more
than once in a term and
are summed over freely subject to the following rules before summations:
\begin{enumerate}
\item Each distinct replica index appears an even number of times~\cite{book}.
\item ${\rm O}(M)$ vector indices appear twice each, and only when the
corresponding replica
indices are the same, due to the ${\rm O}(M)$ symmetry of the randomness.
\item Repetition of a time ``index'' corresponds to interaction of spins,
which must be
local in time and accordingly can be expanded as terms with times set equal
plus the same with
additional derivatives; it occurs when the corresponding replica indices
are the same, and only then.
\end{enumerate}
\item The action should be invariant under space and time translations, and
under space
and time inversions (under which $Q$ is invariant).
\end{enumerate}
(The rules for the indices may be best appreciated diagrammatically or from the
microscopic approach in Appendix~\ref{derivation}.)

This procedure yields the Landau functional (recall that we are using the
Einstein
summation convention for the ${\rm O}(M)$ vector indices):
\begin{eqnarray}
{\cal A} = && \frac{1}{t} \int d^d x \left\{ \frac{1}{\kappa} \int d\tau \sum_a
\left. \left[ \frac{\partial}{\partial\tau_1}
\frac{\partial}{\partial \tau_2} + r \right] Q^{aa}_{\mu\mu} (x , \tau_1 ,
\tau_2 )
\right|_{\tau_1=\tau_2=\tau} + \frac{1}{2} \int  d \tau_1
d \tau_2 \sum_{a,b} \left[ \nabla Q_{\mu\nu}^{ab} (x, \tau_1, \tau_2 )
\right]^2 \right. \nonumber \\
&&~~~~~~~~~~~~~~~ - \frac{\kappa}{3} \int  d \tau_1 d \tau_2 d \tau_3
\sum_{a,b,c} Q^{ab}_{\mu\nu} (x, \tau_1 , \tau_2 ) Q^{bc}_{\nu\rho}
(x, \tau_2 , \tau_3 ) Q^{ca}_{\rho\mu}
(x, \tau_3 , \tau_1 ) \nonumber \\
&&~~~~~~~~~~~\left. + \frac{1}{2} \int  d \tau \sum_a \left[ u~
Q^{aa}_{\mu\nu} ( x, \tau , \tau) Q^{aa}_{\mu\nu} ( x, \tau , \tau)
+v~ Q^{aa}_{\mu\mu} ( x, \tau , \tau) Q^{aa}_{\nu\nu} ( x, \tau , \tau)
\right]\right\} \nonumber \\
&&~~~~~~~~~~~~~~~~~~ - \frac{1}{2t^2} \int d^d x \int  d \tau_1 d \tau_2
\sum_{a,b}
Q_{\mu\mu}^{aa}  (x, \tau_1 , \tau_1 ) Q_{\nu\nu}^{bb} ( x, \tau_2 ,
\tau_2 ) ~~~+~~~ \cdots~.
\label{landau}
\end{eqnarray}
There are four terms in ${\cal A}$ whose co-efficients contain products of
powers of only
two coupling constants, $\kappa$ and $t$; this form can be reached without
loss of generality by
suitably rescaling the space and time co-ordinates. The reasons for our
rather peculiar
choices for these couplings will only become evident when the structure of
perturbation theory
is discussed.
We have only retained the terms which
our later power counting will tell us are relevant or marginal in high space
dimension
$d$, together with the leading irrelevant term. The exception to this
statement is a
quadratic term
\begin{equation}
\int d^d x d \tau_1
d \tau_2 \sum_{a,b} \left[ Q_{\mu\nu}^{ab} (x, \tau_1, \tau_2 )
\right]^2
\label{mass}
\end{equation}
which appears to be highly relevant in all $d$.
However, it is a ``redundant'' operator as it can be removed by a further
transformation
$Q \rightarrow Q - C \delta^{ab} \delta_{\mu\nu} \delta ( \tau_1 - \tau_2
)$ for a
suitable choice of $C$. This relies on the presence of the cubic term with
coefficient
$\kappa /t$, and on the implicit $n \rightarrow 0$ (replica) limit to
eliminate the
contribution of the last $1/t^2$ term. The net effect is that if $r$ is
redefined to
absorb a constant, the action has the form given in (\ref{landau}). We will
see that this
choice of definition of $Q$ to eliminate the term in (\ref{mass}) also
makes $\langle
\langle Q \rangle \rangle = 0$ at $\omega =0$ and $g=g_c$. This leaves $r$,
the coupling
in the term linear in $Q$, as the parameter expected to drive the system
through its
transition.

The possibility of eliminating a quadratic term by such a shift in the
field variable
arises generally in field theories containing a cubic term. The simplest
example is for a
theory with a single scalar field $\phi$, which arises physically in
the Yang-Lee~\cite{yang} edge of
a classical Ising-like ferromagnet in an imaginary magnetic field. The
field induces an
imaginary expectation of $\phi$, which when eliminated by redefining $\phi$
by a shift,
generates, due to the real $\phi^4$ coupling in the Ising system, an
imaginary $\phi^3$
term in the new $\phi$. In the Landau theory of this critical point, due to
M.E. Fisher~\cite{mef},
it is again convenient to shift away the quadratic term and leave a linear
term in $\phi$
in the action. The coefficient, $r$, of this term controls the distance
from the critical
point, where power law correlations of $\phi$ appear. Since $r$ appears in
the same form
as an (additional) external field, there is only a single scaling field
($\phi$) in the
critical theory~\cite{mef,jlc}, and there is a scaling relation between the
exponent $\eta$ governing
correlations of $\phi$ at the critical point, and the correlation length
exponent $\nu$.
In our spin glass model, one might similarly expect an exponent relation,
since the
operator $Q^{aa}_{\mu\mu} ( x, \tau , \tau )$ is the ``thermal'' operator whose
coefficient in ${\cal A}$ drives one across the transition, while
$Q_{\mu\nu}^{ab} (x,
\tau_1 , \tau_2 )$ is the basic order parameter field. However due to the
``trace'' over
replica, $O(M)$ vector, and imaginary time indices in the thermal operator,
it is
far from clear that these two will in fact have the same scaling dimension
in general.
We will address this point further in Section~\ref{scalinghypoth}. However
at mean field level, {\em
i.e.\/} at the Gaussian fixed point, such a relation holds and we will find
$\nu = 1/4$,
the same value as for the Yang-Lee problem in mean field, this being due in
both cases to
the linear plus cubic form of the action, in contrast to the value $\nu =
1/2$ found in
most mean field theories. We note that for the classical spin glass, though
the leading coupling
is cubic, the replica diagonal components $Q^{aa}$ are not critical and are
omitted from the Landau
theory, so the same mechanism does not apply, and $\nu=1/2$ in mean field
theory~\cite{book}.

A few last remarks on the action ${\cal A}$:
the terms with coefficients $u/t$ and $v/t$
arise from quartic couplings of the spins within a single replica (see
Appendix~\ref{derivation}).
They are the only terms retained that break replica ${\rm O}(n)$ symmetry
to $S_n$, the
permutational symmetry. If they were omitted, the action would describe
randomly coupled
simple harmonic oscillators which is definitely an unstable system in
finite dimensions,
and so anharmonic terms (and hence $u$ and $v$) are expected to be necessary
for
stability. The last term with coefficient $1/t^2$ represents randomness in $r$
(with variance $=+1$) which could originate from randomness in $g$ (see
Appendix~\ref{derivation}) but is also generated by the random $J_{ij}$'s.
It turns out to play a
central role.

\subsection{Phases of the model}
\label{meanfieldtheory}
In this subsection we will solve the action ${\cal A}$ in the mean field,
or more
accurately, tree approximation, {\em i.e.\/} without any momentum loop
integrations for
fluctuations. For the infinite range model, where spatial dependence of $Q$
can be dropped
to obtain either thermodynamic quantities as averages over the whole
system, or onsite
correlation functions, this is exact and constitutes a simpler rederivation
of results
obtained earlier~\cite{letter}. For the short-range finite dimensional
quantum models, as in the
classical case~\cite{book}, mean-field theory should be a useful starting
point towards
understanding
the overall phase diagram and properties of the phases. For the critical
properties of the quantum
transition, the mean-field theory is an
approximation whose validity as an attractive weak-coupling fixed point
under renormalization group
in sufficiently high dimensions will be examined in the next section. Here
we will
concentrate on the correlators defined in Section~\ref{observables}, and
study first
in Section~\ref{paracrit} the paramagnetic
phase and critical point. We will then in Section~\ref{spinglassphase}
study the spin glass phase
and the appearance of replica symmetry breaking at finite temperatures.
The properties of these phases in a longitudinal magnetic field will be
discussed in
Section~\ref{longitudinalfield} and in a field coupling to the conserved
angular momentum (defined
only for $M > 1$) in Section~\ref{conservefield}.

\subsubsection{Quantum paramagnet and the critical point}
\label{paracrit}
The saddle-point and perturbative analysis are most conveniently performed
in momentum ($k$) and frequency ($\omega$) space.
We will work at a finite, but small, temperature $T= 1/\beta$, and $\omega$
will
therefore take values at the discrete Matsubara frequencies. The
normalization of the
Fourier transform is set by
\begin{equation}
Q ( k, \omega_1 , \omega_2 ) = \int d^d x \int_{0}^{\beta} d \tau_1 d
\tau_2 Q(x, \tau_1 , \tau_2 )
e^{i (k x - \omega_1 \tau_1 - \omega_2 \tau_2 )}.
\end{equation}
In these Fourier transformed variables
we expect the saddle point-value of $Q$ to obey the following ansatz in the
paramagnet
and at the critical point
\begin{equation}
Q^{ab}_{\mu\nu}(k, \omega_1 , \omega_2 ) = \beta \delta_{\mu\nu} \delta^{ab}
(2 \pi)^{d} \delta^d (k) \delta_{\omega_1 + \omega_2,0} D(\omega_1) .
\label{ansatzpara}
\end{equation}
The momentum and frequency structure of the right hand side follows from
the Fourier
transform of (\ref{defD}). An explicit factor of $\beta$ has been inserted
to make $D(\omega )$
finite in the zero temperature limit. The
structure in the $O(M)$ spin space follows from spin rotation invariance,
while the
replica-diagonal structure follows from the absence of a static moment in
the paramagnetic phase.
Inserting this into ${\cal A}$ in (\ref{landau}), we obtain for the free
energy density ${\cal F}/n$
(as usual, ${\cal F}/n$ represents the physical disorder averaged free energy)
\begin{equation}
\frac{{\cal F}}{n} = \frac{M}{\beta t} \sum_{\omega} \left[ \frac{\omega^2
+ r}{\kappa}
D(\omega) - \frac{\kappa}{3} D^3 ( \omega ) \right] + M \frac{u+Mv}{2t} \left[
\frac{1}{\beta} \sum_{\omega} D (\omega ) \right]^2 .
\label{calf}
\end{equation}
The contribution from the last $1/t^2$ term in ${\cal A}$ vanishes in the
replica limit $n \rightarrow 0$ and is therefore absent. The stationary
point with
respect to variations in $D(\omega )$ gives us the result
\begin{equation}
D( \omega ) = - \frac{1}{\kappa} ( \omega^2 + \tilde{r} )^{1/2}
\label{resd}
\end{equation}
where $\tilde{r}$ is given implicitly by
\begin{equation}
\tilde{r} = r - (u + M v) \frac{1}{\beta} \sum_{\omega} ( \omega^2 +
\tilde{r} )^{1/2}.
\label{tilder}
\end{equation}
The sign of $D(\omega )$ is determined by the fact that the Fourier transform
$D(\tau )$
is positive. This solution for $D(\omega)$ is well-defined for $\tilde{r}
\geq 0$, while
no sensible paramagnetic solution exists for $\tilde{r} < 0$, suggesting
that the
critical line in the $r,T$ plane between the paramagnetic and spin glass phases
is $\tilde{r} = 0$.
The local density of excitations $\chi^{\prime\prime} ( \omega )$ can be
obtained
by analytic continuation of $D(\omega )$ to real frequencies and is therefore
\begin{equation}
\chi^{\prime\prime} ( \omega ) = \mbox{sgn($\omega$)}\frac{ (\omega^2 -
\tilde{r})^{1/2}}{\kappa} \theta( |\omega| - \sqrt{\tilde{r}} ).
\label{chi}
\end{equation}
There is a gap, $\sqrt{\tilde{r}}$, in the spectral density which vanishes
at the
critical point $\tilde{r} = 0$. This gap is expected to be filled in at
finite temperatures
by loop corrections involving inelastic effects; in addition, Griffiths
effects (see
Appendix~\ref{griffiths}) will lead to sub-gap absorption at both zero and
finite temperatures.

{}From (\ref{tilder}) we determine that the critical point $\tilde{r} = 0$
occurs when
\begin{equation}
r = r_c (T) \equiv \frac{u+Mv}{\beta} \sum_{\omega} | \omega |.
\end{equation}
The frequency summation is obviously divergent, and the result will depend
upon the nature
of the ultraviolet cutoff. However the temperature dependence of the result
is entirely in the
subleading term, which turns out to be cutoff-independent (provided the
cutoff is smooth
on the scale of $T$). The summation can be evaluated by the Poisson
summation formula, which yields:
\begin{equation}
r_c (T) = r_c  - (u+Mv) \frac{\pi T^2}{3}
\label{defrct}
\end{equation}
where for a high-frequency cutoff around $\Lambda_{\omega} \sim g$
\begin{equation}
r_c  \equiv r_c (0) = \frac{(u + Mv) \Lambda_{\omega}^2}{2 \pi}.
\end{equation}
The line between the paramagnetic and spin-glass phases is shown in
Fig~\ref{phasediag}
in the $r,T$
plane.

We now turn to determining the behavior of the `gap' parameter $\tilde{r}$
close to $r=r_c$ and at finite $T$, by solving (\ref{tilder}). First we
evaluate the frequency
summation, using the following identity.
\begin{eqnarray}
\frac{1}{\beta} \sum_{\omega} (\omega^2 + \Delta^2)^{1/2} &=&
\frac{\Lambda_{\omega}^2}{2 \pi} + \frac{\Delta^2}{2 \pi} \log ( c_1
\Lambda_{\omega}/\Delta ) + {\cal O} (e^{-\Delta/T})
{}~~~~~~~~~~~~~~~~~~~~~~~~~\mbox{for $\Delta \gg
T$} \nonumber \\
&=& \frac{\Lambda_{\omega}^2}{2 \pi} - \frac{\pi T^2}{3} + T \Delta +
\frac{\Delta^2}{2
\pi} \log ( c_2 \Lambda_{\omega}/T) + {\cal O} (\Delta^3 /T)
{}~~~~~~~\mbox{for $\Delta \ll T$}
\end{eqnarray}
where $c_1, c_2$ are constants of order unity.
Then, upon examining the solution of (\ref{tilder}), we find that there are
three different regimes in the paramagnet phase,
labeled as I, II and III in Fig~\ref{phasediag}. The conditions defining
the regimes,
and the corresponding results for $\tilde{r}$ are given below:
\begin{eqnarray}
&&~~~~~\frac{4 \pi}{(u + Mv)} \frac{ (r-r_c)}{ \log(\Lambda_{\omega}^2 /
(r-r_c ))} ~~~~~~ \mbox{Regime I, $(r-r_c)^{1/2} \gg T$} \nonumber \\
\tilde{r} =&&~~~~~\frac{2 \pi^2 T^2}{ 3\log(\Lambda_{\omega} /
T)}~~~~~~~~~~~~~~~~~~~~~~~~
\mbox{Regime II, $T \gg |r-r_c|^{1/2}$} \nonumber \\
&&~~~~~\frac{(r-r_c (T))^2}{T^2} ~~~~~~~~~~~~~~~~~~~~~~~~
\mbox{Regime III, $(r-r_c (T))^{1/2} \ll T$}.
\label{logr}
\end{eqnarray}
Regime I is that of the quantum paramagnet, where the properties are those
of the
quantum-disordered ground state, and thermal effects
are not important. Regime II is `quantum-critical': here the system behaves
as if it is at the
critical point $r=r_c$, and the properties reflect those of the critical
ground state and
its excitations; it is the analog of the quantum-critical region discussed
in Refs~\cite{chn,csy}
for quantum rotors in the absence of disorder.
Finally, in regime III, close enough to the finite-temperature phase
transition, classical effects
take over completely, and the behavior is that of the usual
finite-temperature spin-glass/paramagnet
transition in the classical model~\cite{book}. Regime III also extends into the
spin glass phase,
although here we have only obtained results for its paramagnetic portion.
Notice that
in regime III, $\tilde{r}$ now depends upon the {\em square} of the
distance from the transition
$r-r_c(T)$: as will become clear from later results, $\tilde{r}^{1/4}$
plays the role
of an inverse correlation length, so this is
just what is needed to transform the quantum model with $\nu=1/4$ to the
classical model with
$\nu=1/2$ in mean-field theory. The results (\ref{chi}) and (\ref{logr})
for the local spectral weight and the
asymptotic form of the gap, including the logarithmic correction are seen to be
identical to those obtained in the infinite range model by different methods
earlier~\cite{miller,letter}, as quoted in Section~\ref{infiniterange}
above (there
was a factor of 2 error in the result for regime II in (\ref{logr}) in
Ref~\cite{letter}).

The asymptotic form of the free energy density, ${\cal F}$ can
be obtained from (\ref{calf}), (\ref{resd}) and (\ref{logr}). At zero
temperature
this yields
\begin{eqnarray}
\frac{{\cal F}(T=0)}{nM} = &&\frac{1}{t\kappa^2} \left\{
-\frac{\Lambda_{\omega}^4}{\pi}
\left( \frac{1}{6} + \frac{u + Mv}{8 \pi} \right) - (r - r_c )
\frac{\Lambda_{\omega}^2 }{2
\pi } - (r-r_c )^2 \frac{1}{2(u + Mv )  }\right.  \nonumber \\
&&~~~~~~~~~~~~~~~~~~~\left. + \frac{(r-r_c)^2}{\log ( \Lambda_{\omega}^2 /
(r - r_c
))}~ \frac{2\pi}{  (u + Mv)^2} + \ldots \right\}.
\label{resfreepara}
\end{eqnarray}
Note that the first three terms in ${\cal F}$ involve only integer powers
of $r-r_c$.
This does not immediately imply that these terms form a smooth background
through the
transition, as there could be discontinuities in the coefficients of
$r-r_c$: a computation on
the spin glass side is required to determine this. Intuitively, we might
guess that the
first two cutoff-dependent terms will be analytic through the transition,
while the
coefficient of the $(r-r_c)^2 $ term may have a discontinuity---we will see
that even
this discontinuity does not appear. The last term has a manifest
logarithmic singularity
at $r=r_c$: we will relate this singularity to marginal operators in
Section~\ref{renormgroup}.

We also examined the temperature dependence of the free energy in regime II
(Fig~\ref{phasediag}),
above
the quantum critical point $r = r_c$, and found
\begin{equation}
\frac{{\cal F}(r=r_c, T)-{\cal F}(r=r_c, T=0)}{nM} = - \frac{4 \pi^3
T^4}{45 t \kappa^2} \left(
1 + {\cal O}(1/\log^{1/2} (\Lambda_{\omega} / T))  \right) + \cdots .
\label{freeent}
\end{equation}
This predicts a low temperature specific heat $\sim T^3$. A scaling
interpretation of the
power of $T$ will be given later in Section~\ref{renormgroup}.

We now turn to a tree-level determination of the correlation functions that
were introduced in
Section~\ref{observables}. To do this
we must expand $Q$ about its saddle-point
value \begin{equation}
Q^{ab}_{\mu\nu}(k, \omega_1 , \omega_2 ) = \beta\delta_{\mu\nu} \delta^{ab}
(2 \pi)^{d} \delta^d (k) \delta_{\omega_1 + \omega_2,0} D(\omega_1)
 + \tilde{Q}^{ab}_{\mu\nu}(k, \omega_1 , \omega_2 )
\end{equation}
and evaluate correlators of $\tilde{Q}$. Expanding ${\cal A}$ to order
$\tilde{Q}^2$
we can obtain the propagator of the $\tilde{Q}$ field. It is easy to see
that when $a\neq
b$, this propagator is in fact $1/M$ times the $G$ correlator (Eqn
(\ref{defgd2}))
\begin{equation}
G ( k, \omega_1 , \omega_2 ) = \frac{Mt}{k^2 + \sqrt{\omega_1^2 + \tilde{r}}
+ \sqrt{\omega_2^2 + \tilde{r}}}.
\label{propagator}
\end{equation}
Note that this propagator has a factor $t$ in the numerator and is
independent of $\kappa$---the
factors of $\kappa$ were placed judiciously in ${\cal A}$ to achieve this.
{}From the form of
(\ref{propagator}) we can deduce that at the critical point $\tilde{r} =
0$, $|\omega | \sim
k^2$, so the dynamic exponent $z=2$, while for $\tilde{r} \neq 0$ there is
a length scale $\xi
\sim \tilde{r}^{-1/4}$ so that the exponent defined by $\xi \sim
(r-r_c)^{-\nu}$ in regime I is
$\nu = 1/4$.

It is useful for the subsequent considerations to develop
a diagrammatic representation of the $\tilde{Q}$ propagator and the
interactions in ${\cal
A}$. As $Q$ is a matrix field, we will use a double-line representation for
the $Q$
propagator, $G$ (Fig~\ref{vertices}a ). Each line can be considered to be
one of the $S$ field
components of the composite $Q$ (Eqn (\ref{composite})), and carries with it a
replica index, spin index, and frequency. The momentum is however
carried by the pair of lines. The $u/t$ and $v/t$ terms now become
two-point interactions
(Fig~\ref{vertices}b) at which frequency is transferred between the four
lines.  The
$\kappa/t$ interaction is the three-point vertex shown in
Fig~\ref{vertices}c, while the
$1/t^2$ term is in Fig~\ref{vertices}d :  no frequency is exchanged between
the lines in
the $\kappa/t$ and $1/t^2$ vertices.

The connected Green's function is given by the sum of all diagrams with
repeated two-point $u$ and $v$ interactions. Such diagrams can be easily
summed and yield
\begin{eqnarray}
G^c_{\mu\nu\rho\sigma} (k , \omega_1 , \omega_2, \omega_3)
= &&- \frac{1}{M^2 t} G (k, \omega_1 , \omega_2 ) G ( k, \omega_3 , \omega_4 )
 \left[\frac{u ( \delta_{\mu\rho} \delta_{\nu\sigma} + \delta_{\mu\sigma}
\delta_{\nu\rho}
)/2}{1 + u L ( \omega_1 + \omega_2 , k, \tilde{r} )} \right.\nonumber \\
&&~~~~~~~\left. + \frac{v  \delta_{\mu\nu} \delta_{\rho\sigma}}{(1 + u L
( \omega_1 + \omega_2 , k, \tilde{r} ))(1 + (u+Mv)L(\omega_1 + \omega_2, k,
\tilde{r}))}
\right]
\label{resgc}
\end{eqnarray}
where the three frequencies $\omega_1$, $\omega_2$, $\omega_3$ arise from
the Fourier
transform of the three time arguments in (\ref{defgc}) and  $\omega_1 +
\omega_2 + \omega_3 +
\omega_4 = 0$, and
\begin{equation}
L( \omega , k, \tilde{r} ) \equiv \frac{1}{\beta} \sum_{\Omega}
\frac{1}{k^2 + \sqrt{\Omega^2
+ \tilde{r}} + \sqrt{(\Omega-\omega)^2 + \tilde{r}}} \approx \frac{1}{2 \pi}
\log \left( \frac{\Lambda_{\omega}}{\mbox{max $(k^2 , |\omega|,
\sqrt{\tilde{r}}, T)$}} \right)
\label{defL}
\end{equation}
is the frequency integral that appears between two two-point vertices (we
have assumed
that we are not in the classical regime III). Note that the
connected Green's function is proportional to the quantum mechanical
interactions
$u$, $v$ as it should be. Further note that $G^{c} \sim t$ as the two $G$'s
carry a
factor of $t$.

Finally we can also evaluate the disconnected Green's function $G^{d}$. This is
obtained by attaching $G^c$ and $G$ on the ends of a $1/t^2$ vertex
(Fig~\ref{vertices}d), and carrying out the intermediate frequency integral
(it is not necessary to
include graphs with repeated $1/t^2$ insertions as they all vanish in the
replica limit
$n\rightarrow 0$); this gives
\begin{equation} G^{d} (k, \omega_1 , \omega_2 ) =  \frac{1}{t^2} \frac{G
(k, \omega_1, -\omega_1)
G (k, \omega_2 , -\omega_2 ) }{( 1 + (u + Mv) L (0, k, \tilde{r}))^2}.
\label{resgd1}
\end{equation}
Note that the factor of $1/t^2$ with the vertex  will cancel against the
two $t$'s in the
$G$'s, and $G^{d}$ is independent of $t$.
It is seen that, even neglecting the denominators that contain $L$,
$G^c$ and $G^d$ both vary as $G^2$, so are more strongly divergent at long
wavelengths
than $G$.

The results for the correlation functions now allow us to define the
correlation exponent
$\eta$. The basic correlation function is $G$, and we define $\eta$ by
\begin{equation}
G(k, 0, 0) \sim k^{-2 + \eta}
\label{defeta}
\end{equation}
so that $\eta=0$ in mean field theory, as is conventional. In real space,
$G$ then decays
as $x^{-(d+2z-2+\eta)}$. Our definition, though conventional in its
relation to mean field
theory, differs from that used in two recent papers~\cite{guo,rieger}.
Their $\eta$,
which we call $\eta^{\prime}$, is related to ours by $\eta^{\prime} = \eta +
z$.
We can also define another exponent for $G^d$, in analogy with the random
field Ising
model, by
\begin{equation}
G^d (k, 0, 0) \sim k^{-4 + \bar{\eta}}
\label{defbareta}
\end{equation}
so that $\bar{\eta}$ is also zero in the Gaussian approximation.
In real space $G^d \sim x^{-(d+2z-4+\bar{\eta})}$.
Discussion of these
exponents will be continued in Sections~\ref{renormgroup}
and~\ref{scalinghypoth}.

\subsubsection{Spin glass phase}
\label{spinglassphase}
We will now look at the mean-field behavior of ${\cal A}$ for $r < r_c (T)$
where
spin-glass order appears. This phase was studied in Ref~\cite{letter} for
the infinite-range
model only at $M=\infty$, and a replica-symmetric solution for the $Q$
order parameter was
obtained at all temperatures. The absence of replica-symmetry breaking was
not surprising as the classical
model is also known to preserve replica symmetry at
$M=\infty$~\cite{thouless}. The present mean-field
theory is being carried out at finite $M$, and so one expects, first, that
at finite temperature
replica symmetry breaking should occur since the system maps onto the
classical system, and second
that at zero temperature the ordered ground state resembles that of a
classical system, However,
this leaves open the question how replica symmetry breaking behaves as $T
\rightarrow 0$. This
turns out to be quite subtle already in the classical spin
glass~\cite{youngbinder,pat,temesvari},
with which we will compare and contrast our results after they have been
described.

We will begin this section with a mean-field treatment of ${\cal A}$ (Eqn
(\ref{landau})), which
yields a replica-symmetric solution for the spin-glass phase at all
temperatures. We will then consider
the consequences of adding higher-order terms to ${\cal A}$, terms which
are formally irrelevant
under the subsequent renormalization group analysis of
Section~\ref{renormgroup}; all such terms will
continue to be innocuous at zero temperature, but one of them becomes
dangerous at any finite
temperature and leads to replica symmetry breaking, whose strength is
proportional to temperature.
The mechanism of the replica symmetry breaking at low temperatures will be
discussed
within the framework of our Landau theory.

The generalization of the saddle-point
ansatz (\ref{ansatzpara}) to the spin glass phase is
\begin{eqnarray}
Q^{ab}_{\mu\nu} ( k, \omega_1 , \omega_2 ) &=& (2 \pi)^d \delta^{d} ( k )
\delta_{\mu\nu}
\left( \beta D( \omega_1 ) \delta_{\omega_1 + \omega_2,0} \delta^{ab}
+ \beta^2 \delta_{\omega_1,0} \delta_{\omega_2 , 0} q^{ab}\right) \nonumber \\
D ( \omega ) &\equiv& \bar{D} ( \omega ) + \beta \tilde{q} \delta_{\omega , 0}
{}.
\label{ansatzsg}
\end{eqnarray}
We have introduced the parameters $q^{ab}$, $\tilde{q}$ and $\bar{D} (\omega)$,
which have to be varied to determine the stationary point of the free energy.
As before, factors of $\beta$ have been judiciously placed to ensure that
all these parameters
are finite in the limit $\beta \rightarrow \infty$.
Note that the replica off-diagonal components of $Q$ at the saddle-point
are non-zero only when
both frequencies are zero: this is because they represent averages of the type
$\left[\left\langle S(\tau_1 ) \right\rangle\left\langle S ( \tau_2 )
\right\rangle \right]$
which must be time independent by time translational invariance of the
quantum mechanical
averages. Without loss of generality we may assume that the diagonal
components of
$q^{ab}$ are zero, and that $\bar{D}(\omega = 0 ) =0$, as both of these can
be absorbed into
$\tilde{q}$. We expect a solution in which $\bar{D}(\omega)$ is finite and
continuous (except
perhaps at $\omega=0$) as $\beta \rightarrow \infty$, in which case
(\ref{defqea}) implies that
at zero temperature $\tilde{q}$ is just the Edwards Anderson order
parameter $q_{EA}$;
the equality between $q_{EA}$ and $\tilde{q}$ does not hold at non-zero
temperatures,
although we always have $q_{EA} = \mbox{max}_{a\neq b} q^{ab}$~\cite{book}.
We now insert this ansatz into ${\cal A}$ and find for the free energy density
${\cal F}$
\begin{eqnarray}
\frac{{\cal F}}{Mn} = && \frac{1}{\beta \kappa t} \sum_{\omega} (\omega^2 +
r) \bar{D} ( \omega
)
 + \frac{r}{\kappa t} \tilde{q} - \frac{\kappa}{3 \beta t} \sum_{\omega}
\bar{D}^3 ( \omega ) +
\frac{(u + Mv)}{2t}
\left(\tilde{q} +  \frac{1}{\beta} \sum_{\omega} \bar{D}(\omega ) \right)^2
\nonumber \\
&&~~~~~~~~~~~~~~~~~~~~~~~~~~~~~~~- \frac{\kappa \beta^2}{3t} \left(
\tilde{q}^3 + 3 \tilde{q} \frac{\mbox{Tr} q^2}{n} + \frac{\mbox{Tr} q^3}{n}
\right) .
\end{eqnarray}
All factors of $\beta$ have been written explicitly and all other variables
are expected to
be finite in the limit $\beta \rightarrow \infty$. Let us now examine this
limit in ${\cal F}$. In the
first four terms, we only have factors of $1/\beta$ associated with frequency
summations, and the combinations should be finite as $\beta \rightarrow
\infty$. The remaining
terms all arise from the static contributions to the $Q^3$ term in ${\cal
A}$, and appear to
have a dangerous divergence $\sim \beta^2$ as $\beta \rightarrow \infty$.
We will determine the saddle point of ${\cal F}$ for finite $\beta$ below,
and find that
in the replica limit of $n \rightarrow 0$, all the terms of order $\beta^2$
(and also of order
$\beta$) in fact cancel with each other at the saddle point, yielding a
finite zero temperature free
energy density.

We now find the saddle-point of ${\cal F}$ with respect to variations in
$\tilde{q}$, the
function $\bar{D} (\omega)$, and the space of ultrametric matrices $q^{ab}$.
This can be done by a straightforward extension of the classical
methods~\cite{book} and we
simply quote the final results, valid for arbitrary $\beta$
\begin{eqnarray}
\tilde{q} &=& \frac{1}{\beta\kappa} \sum_{\omega} |\omega|  -
\frac{r}{\kappa(u + Mv)}
\nonumber \\
&=& \frac{1}{(u+Mv)\kappa} ( r_c (T) - r) \label{restildeq} \\
q^{ab} &=& \tilde{q} ~~~\mbox{for $a\neq b$} \label{resqab} \\
\bar{D} ( \omega ) &=& - \frac{|\omega|}{\kappa}
\label{resdom}
\end{eqnarray}
where $r_c (T)$ was defined in (\ref{defrct}).
Note that this solution is replica symmetric
as all off-diagonal matrix elements of $q^{ab}$ are equal. The order parameter
$q^{ab}$ has to be positive in the spin-glass phase, which is therefore
restricted to $r < r_c (T)$.
At zero
temperature, the properties of $D$ are a little more transparent in the
time domain,
\begin{equation}
D ( \tau ) = \tilde{q} + \frac{1}{\pi \kappa \tau^2}~~~\mbox{at $T=0$},
\end{equation}
where it decays to a finite, positive value ($=\tilde{q}=q_{EA}$) at large
time.
Note also that the
classical definition~\cite{book} $q_{EA} = \mbox{ max}_{a\neq b} q^{ab}$, and
(\ref{resqab}) give the same value of $q_{EA}$; this equality between the
two approaches to
$q_{EA}$ is also easily seen to lead to a cancellation of all terms of
order $\beta^2 $ in
${\cal F}$. The $1/\tau^2$ decay in $D(\tau )$ also holds at the critical
point $r=r_c , T=0$ where
the decay is to zero. The power-law decay is related to the gaplessness of
the spectral density in
the entire spin glass phase (from (\ref{resdom}) $\chi^{\prime\prime} (
\omega ) = \omega / \kappa
$).

The results (\ref{restildeq}--\ref{resdom}) are actually valid in both
regimes III and IV
of Fig~\ref{phasediag} in the spin-glass phase. As a result, the order
parameter exponent
$\beta =1$ in both the classial and quantum transitions. Crossovers between
regimes III and IV will
presumably appear upon considering fluctuations about the present
mean-field theory.

 We also quote the
result for the free energy density in the spin glass phase at zero temperature:
\begin{equation}
\frac{{\cal F} (T=0)}{nM} = -\frac{\Lambda_{\omega}^4}{\pi \kappa^2 t}
\left( \frac{1}{6} + \frac{u +
Mv}{8 \pi} \right) - (r - r_c ) \frac{\Lambda_{\omega}^2 }{2 \pi \kappa^2
t} - (r-r_c )^2
\frac{1}{2(u + Mv ) \kappa^2 t} .
\label{resfsg}
\end{equation}
This differs from the expression (\ref{resfreepara}) for the paramagnetic
phase only
by the absence of the term with a logarithmic singularity. The terms in
(\ref{resfsg}) thus {\em
do\/} constitute an analytic background to the singular part.

\paragraph*{Replica symmetry breaking:} It is known from the Landau theory
of the classical spin
glass that replica symmetry breaking does not appear until terms of order
$Q^{4}$ have been included
in the action~\cite{book}. So to address the stability of the replica symmetric
mean-field solution (\ref{restildeq})-(\ref{resdom}), we have to extend
${\calA}$ to include at least such
terms. Upon examining these terms, a second obstacle immediately confronts
us: one of the quartic
terms has 4 time integrals, so the contribution of the static moments to
the free energy appears to
diverge as
$\beta^3$ in the zero temperature limit. It is also easy to see that this
obstacle gets worse
with terms of higher order, which appear to contribute even higher powers
of $\beta$ to the free
energy. It might therefore appear that it is necessary to resum these
strongly divergent terms and
the whole Landau theory framework is breaking down in the spin glass phase.
We now argue that this
is not the case, and that enough factors of $\beta$ cancel out to yield
finite results for
$\tilde{q}$, $q^{ab}$ and $\bar{D}(\omega)$ in the $\beta \rightarrow
\infty$ limit.
One can proceed by the usual Landau theory framework to develop an
expansion for all physical
quantities in powers of $(r_c (T) - r)$, and all dangerous factors of
$\beta$ will cancel out order
by order. We illustrate this cancellation explicitly for the quartic terms.
We consider 3 of the many
quartic terms which can be added to (\ref{landau}) (we drop all vector
indices $\mu,\nu$ and treat
only the case
$M=1$, as the vector nature plays no role in the following):
\begin{eqnarray}
{\cal A} = \ldots - && \frac{1}{6t} \int d^d x \left\{ y_1 \int d\tau_1
d\tau_2 \sum_{a,b} \left[ Q^{ab}
(x, \tau_1 , \tau_2 ) \right]^4 \right. \nonumber \\
+ && y_2 \int d\tau_1 d\tau_2 d\tau_3 \sum_{a,b,c} [Q^{ab} (x, \tau_1 ,
\tau_2 )]^2
[Q^{bc} (x, \tau_2 , \tau_3 )]^2 \nonumber \\
+ &&\left. y_3 \int d\tau_1 d \tau_2 d\tau_3 d\tau_4 \sum_{a,b,c,d}
Q^{ab} (x, \tau_1 , \tau_2 ) Q^{bc} (x, \tau_2 , \tau_3 ) Q^{cd} (x, \tau_3
, \tau_4 ) Q^{da} (x, \tau_4 ,
\tau_1 ) \right\}
\label{ycouplings}
\end{eqnarray}
where the initial dots denote the terms already in (\ref{landau}).
Let us now insert the mean-field ansatz (\ref{ansatzsg}) into the extended
action ${\cal A}$. The analysis
for arbitrary ultrametric matrices $q^{ab}$ is rather
complicated---fortunately the
general principles become clear by allowing for a small amount of replica
symmetry breaking. We will
begin by considering the replica symmetric case, and then add a
perturbation to allow for
replica symmetry breaking. For the replica symmetric case $q^{ab} = q$ for
all $a\neq b$, we find for
the free energy density:
\begin{eqnarray}
{\cal F} = && -\frac{n}{t} \left[
\frac{\beta^2 \kappa }{3} (\tilde{q} - q)^2 (\tilde{q} + 2 q) + \frac{\beta
y_1 }{6}
(\tilde{q} - q)(\tilde{q} + q)(\tilde{q}^2 + q^2 ) \right. \nonumber \\
&&~~~~~~~~~~~~~~~+ \left. \frac{\beta^2 y_2}{6} (\tilde{q} - q)^2
(\tilde{q} + q)^2
+ \frac{\beta^3 y_3}{6} (\tilde{q} - q)^3 ( \tilde{q} + 3 q) \right] + \cdots .
\label{quarticstatic}
\end{eqnarray}
We have explicitly written down terms which depend only upon the static
moments; all omitted
terms are explicitly finite in the limit $\beta \rightarrow \infty$ and
play no role in the following
discussion. The key observation that we make from (\ref{quarticstatic}) is
that, term by term, powers
of $\beta$ are always paired with an equal number of powers of $(\tilde{q}
- q)$: a mean-field solution with
$(\tilde{q} - q) \sim 1/\beta$ will therefore give a finite limit as
$\beta\rightarrow 0 $.
That this actually occurs can be
verified by obtaining an explicit expression for $q$ in terms of $\tilde{q}$:
we take the derivative of ${\cal F}$ with respect to $q$ (all terms not
explicitly displayed
in (\ref{quarticstatic}) are independent of $q$), equate it
to zero, and obtain an expansion for $q$ in terms of $\tilde{q}$; this yields
\begin{equation}
q = \tilde{q} + \frac{y_1 \tilde{q}^2 }{3 \beta \kappa} \left( 1 -
\frac{2y_2}{3 \kappa} \tilde{q}
+ \frac{y_3}{3 \kappa^2} \tilde{q}^2 \right),
\label{qmtq}
\end{equation}
with omitted terms being higher order in both $\tilde{q}$ and $1/\beta$.
Here we see an example of our earlier claim that the Landau type expansion
in powers of the
order parameter, $\tilde{q} \sim q$, is well behaved, with no singular
powers of $\beta$ appearing. We also
see that $\tilde{q} - q \sim 1/\beta$ as desired. Note that the most
important contribution to
$\tilde{q} - q$ comes from $y_1$, the term associated with the fewest
powers of $\beta$ in the original
expansion for ${\cal F}$. Terms associated with $y_2$ and $y_3$, had higher
powers of $\beta$
in ${\cal F}$, make contributions $\tilde{q} - q$ which are higher order in
$\tilde{q}$.
This happens because
\begin{equation}
\beta (\tilde{q} - q) \sim \tilde{q}^2 ;
\end{equation}
 so each factor of
$\beta (\tilde{q} - q )$ actually ends up behaving like two powers of the
order parameter.
Thus contrary to naive expectations, it is the terms with the fewest powers
of $\beta$ which are
most important in the Landau expansion.

Let us now add some replica symmetry breaking to the mean-field ansatz
\begin{equation}
q^{ab} = q + \alpha^{ab}~~~~~~~a\neq b
\label{deformq}
\end{equation}
where $\alpha^{ab}$ is a small perturbation along the direction in replica
space where symmetry breaking is
expected. We choose $\alpha^{ab}$ to lie along the `replicon' direction of
de Almeida and Thouless~\cite{AT},
proportional to an eigenvector along which they found an instability of the
replica symmetric solution:
$\alpha^{ab} = \alpha$ for $a,b > 2$, $\alpha^{ab} = \frac{1}{2}(3-n)
\alpha $ for $a=1,2$ and $b > 2$,
$\alpha^{12} = \frac{1}{2} (3-n)(2-n) \alpha$, and $\alpha^{ba} = \alpha^{ab}$.
For this form of $q^{ab}$, we expand the free energy to order $\alpha^2$,
and obtain in addition to the terms in (\ref{quarticstatic})
\begin{equation}
{\cal F} = \ldots  - \frac{6 \beta \alpha^2}{t} \left[  \beta \kappa
(\tilde{q} - q) +  y_1 q^2
+ \frac{\beta y_2 }{3} (\tilde{q}^2 - q^2 ) + \beta^2 y_3 (\tilde{q} - q)^2
\right] + \cdots .
\label{alpha2}
\end{equation}
Again factors of $\beta$ are paired with $(\tilde{q} - q)$ and the
combination will be finite
as $\beta \rightarrow \infty$. There is however an overall prefactor of
$\beta \alpha^2$ which
needs attention. By analogy with the low temperature properties of the
solution to the classical
spin glass,
we may expect~\cite{youngbinder} that the magnitude of the replica symmetry
breaking
will be proportional to $1/\beta$, in which case $\alpha \sim 1/\beta$. The
contribution of the
replica broken component of $q$ to ${\cal F}$ is now seen to be $\sim
1/\beta$, as
is also the case in the classical limit~\cite{youngbinder}. We will verify
these expectations in a
more complete analysis of the replica symmetry breaking below. For now we
note that if
we insert the expansion (\ref{qmtq}) in (\ref{alpha2}) we find the
$\alpha$-dependent contribution
to the free energy
\begin{equation}
{\cal F} = \ldots - 4 \beta \alpha^2 y_1 \tilde{q}^2 /t +\cdots.
\label{falpha}
\end{equation}
The couplings $\kappa$, $y_2$, and $y_3$ have canceled out and there is an
instability towards
replica symmetry breaking driven solely by $y_1$; the analogous result for
the classical model is
well-known~\cite{youngbinder}.

Let us now reiterate the rather simple conclusion to which the above chain
of reasoning has lead us:
the low temperature replica symmetry broken state of the spin glass phase,
can be determined simply
by adding to ${\cal A}$ the single quartic term proportional to $y_1$, and
solving the saddle-point
equations. All corrections from other terms will involve higher-powers of
$\tilde{q}$, and
hence $(r_c - r)$, and no dangerous powers of $\beta$ will appear in this
expansion.
The $q^{ab}$ contribution to ${\cal F}$, including the $y_1$ term, for
general $q^{ab}$
has the form
\begin{equation}
{\cal F} = \ldots - \frac{\kappa \beta^2}{3t} \left( \mbox{Tr} q^3 + 3
\tilde{q} \mbox{Tr} q^2
\right) - \frac{y_1 \beta}{t} \sum_{a\neq b} (q^{ab})^4
+ \cdots .
\end{equation}
It is a straightforward matter~\cite{book} to obtain the optimum
saddle-point of ${\cal F}$
with respect to arbitrary ultrametric matrices $q^{ab}$. Such matrices are
characterized
by a function $q(s)$ on the interval $0 \leq s \leq 1$, which is found to
be of the form~\cite{book}
\begin{equation}
q(s) = \left\{ \begin{array}{cc}
 (s/s_1) q(1) &~~~~~~~~\mbox{for $0 < s < s_1$} \\
q(1)        &~~~~~~~~\mbox{for $s_1 < s < 1$} \end{array} \right.
\label{parisifunction}
\end{equation}
with (expanding to the appropriate order in $\tilde{q}$):
\begin{eqnarray}
q(1) &=& \tilde{q} + \frac{y_1 \tilde{q}^2 }{\beta \kappa} \nonumber \\
s_1 &=& \frac{2 y_1 q(1)}{\beta \kappa} .
\label{parisibreak}
\end{eqnarray}
To leading order in powers of $r - r_c (T)$, we may use the value of
$\tilde{q}$ from
(\ref{restildeq}) in the above. Note that $q(s)$ differs from a constant
over a region of length
$s_1$ which is therefore a measure of the strength of replica symmetry
breaking; as expected this
vanishes as
$1/\beta$ at low temperature.
Alternatively we may compute, from the above and (\ref{restildeq}), the
`broken ergodicity' order parameter
$\Delta_q$~\cite{book}
\begin{eqnarray}
\Delta_q &\equiv& q(1) - \int_0^1 q(s) ds \nonumber \\
&=& \frac{y_1}{(u+Mv)^2 \kappa^3} T (r_c (T) - r)^2;
\label{broken}
\end{eqnarray}
notice the prefactor of $T$ indicating suppression of replica symmetry
breaking at low temperatures.
A renormalization-group
interpretation of the dependencies on $T$ and $r_c - r$ will be given in
Section~\ref{renormgroup}.

The meaning of the statement that replica symmetry breaking disappears as
$T \rightarrow 0$ in the
spin glass phase is as follows. For the infinite range classical spin
glass, it is well known that
Parisi's order parameter function $q(s)$ is related to the
existence of many minima in the energy landscape, separated by infinite
energy barriers in the
ordered phase~\cite{youngbinder,book,mezard}. Thus ``ergodicity is broken''
and (in each realization
of the disorder) configuration space breaks up into disjoint regions,
labeled by $\alpha$, on each
of which a Gibbs weight $e^{- \beta f_{\alpha}}$ can be computed by summing
$e^{-\beta H}$ over all
configurations in the region. The partition function is then $Z =
\sum_{\alpha} e^{- \beta
f_{\alpha}}$ and $P_{\alpha} = e^{-\beta f_{\alpha}}/Z$ is the Gibbs
probability for each of the
``pure phases'' $\alpha$, for a given set of random bonds. Each phase
$\alpha$ has a thermal average
spin $m_i^{\alpha}$ for each site $i$; $q^{\alpha\beta} = (1/N)\sum_i
m_i^{\alpha} m_i^{\beta}$ is the
overlap of different phases, and
\begin{equation}
P_J (q) = \sum_{\alpha\beta} P_{\alpha} P_{\beta} \delta ( q - q^{\alpha\beta})
\end{equation}
is the Gibbs probability of finding overlap equal to $q$. The disorder
average $P(q)$ of $P_J (q)$
is given in terms of Parisi's $q(s)$ by $P(q) = (dq(s)/ds)^{-1}$. A further
key result is that the
free energies $f_{\alpha}$ are independently exponentially distributed
random variables (due to the
randomness in the $J_{ij}$'s): their distribution is
\begin{equation}
P (f ) \propto e^{\rho f}
\end{equation}
where $\rho$ is a function of $T$ and any external fields. Thus $\rho$ is
the inverse spacing of the
low-lying $f_{\alpha}$'s. It turns out that $\rho = \beta s_1$, where $s_1$
is the value of $s$ such
that $q(s) = q(1)$ for $s > s_1$~\cite{mezard}, and that similar results
fold for the distributions
of {\em clusters\/} of pure phases.

We assume that similar results hold for the quantum spin glass (with sums
over states replaced by
traces ). Then our result that $s_1 \rightarrow 0$ as $T \rightarrow 0$
does not mean that many pure
phases do not exist, but rather, since by (\ref{parisibreak}) $\rho
\rightarrow \mbox{constant}$,
that the typical energy differences approach a constant, and so only the
lowest free energy phase is
significant in this limit.
This differs from what has been found in the {\em classical\/} Ising spin
glass, where $q (s)
\rightarrow 1$ as $T \rightarrow 0 $ for each $s$, but $s_1 \rightarrow
1/2$~\cite{pat,temesvari}.

\subsubsection{Phases in a longitudinal field}
\label{longitudinalfield}
We now consider the behavior of the system in a field, $h$,
which couples linearly to the  on-site spin field. In such a field, the
Hamiltonian of the Ising
model (\ref{hamising}) is modified by
\begin{equation}
{\cal H}_I \rightarrow {\cal H}_I - h \sum_i \sigma^z_i ,
\end{equation}
while for the rotor Hamiltonian (\ref{hamrot}) we have
\begin{equation}
{\cal H}_R \rightarrow {\cal H}_R - h \sum_i n_{1i}
\end{equation}
with the field pointing along the $1$ direction. Carrying $h$ through the
derivation
of the effective action in Appendix~\ref{derivation}, we find the following
modification to the
Landau action
\begin{equation}
{\cal A} \rightarrow {\cal A} - \frac{h^2}{2t} \int d^d x d\tau_1 d\tau_2
\sum_{ab}
Q^{ab}_{11} ( x, \tau_1 , \tau_2 )  - \frac{\beta}{2} \chi_{hb} h^2
\end{equation}
where $\chi_{hb}$ is a background, local, contribution to the linear
susceptibility.
We will now extend the above mean-field theory to include the additional
term. It turns out
to be useful to consider the cases $M=1$ and $M> 1$ separately, as there
are significant
differences between them.

\paragraph{$M=1$, the Ising spin glass:} The field term acts like a source
for the off-diagonal
components of $Q$, which are therefore always non-zero. We will first find
the complete mean-field
solution in the absence of $Q^4$ terms, in which case, replica symmetry
will be unbroken and
all the off-diagonal components of $q^{ab}$ will be equal. We will then
proceed to examine
the consequences of a small quartic $y_1$ coupling, and will find that
replica symmetry
breaking occurs at finite temperature for small enough $h$ and $r < r_c
(T)$. The instability line towards
replica symmetry breaking is of course the quantum analog of the well known
de~Almeida-Thouless
(or AT) line~\cite{AT}.

First, the solution at $y_1 = 0$. We insert the ansatz (\ref{ansatzsg})
into ${\cal A}$,
determine stationary points with respect to variations in $\tilde{q}$,
$q^{ab}$, and
$\bar{D} ( \omega )$ and find the following solution, valid everywhere
provided $h\neq 0$:
\begin{eqnarray}
\bar{D} ( \omega ) &=& - \frac{1}{\kappa} (\omega^2 + \Delta^2
)^{1/2}~~~,\omega \neq 0 \nonumber \\
q^{ab} &=& q =  \frac{h^2}{4 \Delta}~~~~~a\neq b \nonumber \\
\tilde{q} &=& \frac{h^2}{4 \Delta} - \frac{\Delta}{\kappa \beta}
\label{solnh}
\end{eqnarray}
where we have as before $\bar{D} ( 0) = 0$, and the parameter $\Delta$ is
determined implicitly by the
equation
\begin{equation}
\Delta^2 = r + (u+v) \left( \frac{\kappa h^2}{4 \Delta} -
\frac{1}{\beta} \sum_{\omega} (
\omega^2 +
\Delta^2 )^{1/2} \right) .
\label{Delta}
\end{equation}
\paragraph*{Small $h$ behavior:}
For small $h$, the solution to these equations depends strongly on the sign
of $r - r_c (T)$,
and we consider the different cases separately.

In the paramagnetic phase $r > r_c (T)$, and we see by comparing
(\ref{Delta}) and (\ref{tilder})
that $\Delta^2 \rightarrow \tilde{r}$ as $h \rightarrow 0$. This value can
be inserted
in (\ref{solnh}) to obtain the low field dependence of $q$ and $\tilde{q}$.
We also expanded the free energy to order $h^4$ and thus obtained the
linear susceptibility,
$\chi_h$
\begin{equation}
\chi_h = \chi_{hb} - \sqrt{\tilde{r}}{\kappa}
\end{equation}
and the non-linear susceptibility $\chi_{nl}$
\begin{equation}
\chi_{nl} = \frac{1}{4 \tilde{r} }\frac{u+v}{1 + (u+v) L(0,0,\tilde{r})}
\end{equation}
where $L$ was defined in (\ref{defL}), and $\tilde{r}$ is given by
(\ref{logr}) for small $r - r_c
(T)$; note that the result for $\chi_{nl}$ is consistent with
(\ref{defchinl}) and
(\ref{resgc}).
As expected, at the transition where $\tilde{r}=0$, $\chi_h$ remains
finite, while $\chi_{nl}$
diverges. The singularity of $\chi_{nl}$ at the $T=0$ quantum transition is
$\chi_{nl} \sim 1/(r-r_c) $, while at the finite $T$ classical transition
we have $\chi_{nl} \sim 1/(r-r_c (T))$.  The
mean-field result for the $T=0$ singularity in $\chi_{nl}$ as $r
\rightarrow r_c$
($\chi_{nl} \sim (r-r_c)^{-1}$)
differs from
that in the infinite range model~\cite{miller}
($\chi_{nl} \sim (r-r_c)^{-1/2}$ up to logarithms)~\cite{private_huse}.

Precisely at the $T=0$ paramagnetic/spin-glass phase boundary, we find the
following
low-field dependencies
\begin{equation}
\tilde{q} = q = \frac{h^{4/3} \log^{1/3} ( \Lambda_{\omega} /
\lambda_{\omega} )}
{(32 \pi \kappa )^{1/3}} .
\end{equation}
where $\lambda_{\omega} = (\kappa h^2)^{1/3} $, while the free energy behaves
as
\begin{equation}
{\cal F}(h) - {\cal F}(h=0) = n \frac{3}{4t} \left( \frac{\pi}{16 \kappa^2
\log (\Lambda_{\omega} /
\lambda_{\omega} )} \right)^{1/3} h^{8/3}
\end{equation}

Above the zero field spin glass phase we find $\Delta = (u+v) \kappa h^2 /
(4 (r_c (T) - r)) +
{\cal O}(h^6)$, which gives from (\ref{solnh})
\begin{eqnarray}
q &=& \frac{1}{(u+v) \kappa} ( r_c (T) - r)  + {\cal O}(h^4) \nonumber \\
\tilde{q} &=& q - \frac{h^2}{4 q \kappa \beta} .
\label{qtildeqh}
\end{eqnarray}
The absence of any $h^2$ term in $q$ leads to the vanishing of the $Q$ field
contribution to $\chi_h$, which is now given only by its background value
\begin{equation}
\chi_h = \chi_{hb}.
\end{equation}
Recall that the classical model~\cite{youngbinder} also has a constant
linear susceptibility
in the spin glass phase. The non-linear susceptibility can be
obtained by expanding the free energy to order
$h^4$, and we get
\begin{equation}
\chi_{nl} = \frac{u+v}{4 (r_c(T) - r)}.
\end{equation}

\paragraph*{Finite $h$ properties:} The main phenomenon at finite $h$ is
the appearance of the AT
surface in the phase diagram; replica symmetry breaking occurs at fields
below this surface
(see Fig~\ref{ATsurface}). Determination of the position of the initial
instability towards
replica symmetry breaking closely parallels the $h=0$ calculation of
Section~\ref{spinglassphase}.
For non-zero $h$, the generalization of (\ref{qmtq}) is
\begin{equation}
q = \tilde{q} + \frac{y_1 \tilde{q}^2}{3 \beta \kappa} + \frac{h^2}{4
\tilde{q} \beta \kappa},
\end{equation}
where we have turned on quartic $y_1$ coupling, but not the unimportant
$y_2$, $y_3$ terms (
this equation is consistent with (\ref{qtildeqh})).
We now perform a replica symmetry breaking deformation of $q$ as in
(\ref{deformq}),
and evaluate the change in the free energy to order $\alpha^2$. This
generalizes (\ref{falpha})
to
\begin{equation}
{\cal F} = \ldots + \frac{6 \beta \alpha^2}{t} \left( \frac{h^2}{4
\tilde{q}} - \frac{2 y_1
\tilde{q}^2}{3} \right)  +\cdots~.
\end{equation}
There is an instability towards replica symmetry breaking only above the
spin-glass region, $r < r_c (T)$, in which case we find from (\ref{qtildeqh})
the following final
result for the position of the AT instability
\begin{equation}
h_{AT}^2 = \frac{8 y_1}{3 \kappa^3 (u+v)^3} \left( r_c (T) - r \right)^3 .
\label{ATres}
\end{equation}
A sketch of $h_{AT}$ as a function of $r$ and $T$ is shown in
Fig~\ref{ATsurface}.
An important feature of the above result is that $h_{AT}$ has a non-zero
limit as $T \rightarrow 0$.
So, even though the strength of the replica symmetry breaking  becomes
vanishingly small
as $T \rightarrow 0$, it requires a finite field strength to restore replica
symmetry even at an infinitesimal $T$. We can quantify this by obtaining an
explicit solution for the
replica symmetry breaking  for $h$ below $h_{AT}$. The computation
parallels that of
Section~\ref{spinglassphase} and that for the classical case~\cite{book}:
the only modification of
the $h=0$ result for the order parameter $q(s)$ in (\ref{parisifunction})
is that there is an
additional plateau at $q(s) = q_0$ for $0 < s < s_0 = s_1 q_0 /q(1)$ with
\begin{equation}
q_0 = \left( \frac{3 h^2}{8 y_1 } \right)^{1/3} .
\end{equation}
As in (\ref{broken}) we can also obtain the `broken ergodicity' order parameter
\begin{equation}
\Delta_q = \frac{ (9 y_1 )^{1/3}}{4 \kappa} T \left(
h_{AT}^{4/3} - h^{4/3} \right)
\label{rsbAT}
\end{equation}
which is clearly non-zero for $h < h_{AT}$. Notice however the prefactor of
$T$, which indicates
the weakness of the replica symmetry breaking.

\paragraph{$M > 1$, quantum rotor spin glass:}
We now have to consider the distinct behavior of the components of $Q$
which are longitudinal
and transverse to the applied field. For a field along the $1$ direction,
the replica
off-diagonal components of the longitudinal $Q_{11}$ are always non-zero,
while those
of the transverse $Q_{\mu \mu}$, $\mu > 1$, do not couple linearly to the
applied field,
and  need not be
non-zero. The Gabay-Toulouse (GT) line~\cite{book} identifies the boundary
along which transverse,
replica off-diagonal components of
$Q$ turn on, and we will determine its position below.

We find it slightly more convenient to determine the GT boundary by
approaching it from below,
where the transverse, replica-off-diagonal components of $Q$ are non-zero.
We generalize the ansatz
(\ref{ansatzsg}) by introducing the longitudinal parameters $q_L^{ab}$,
$\tilde{q}_L$,
and $\bar{D}_L ( \omega )$  for $\mu=1$,
and the transverse parameters
$q^{ab}_T $, $\tilde{q}_T$, and $\bar{D}_T (\omega)$
for $\mu > 1$. As the
GT boundary is present even in the absence of replica symmetry breaking, we
will work with the action
${\cal A}$ with the quartic terms omitted.  In this case $q^{ab}_L = q_L$
for $a \neq b$ and
similarly for $q_T^{ab}$. The generalization of the saddle-point equations
(\ref{solnh}) to the
$M>1$ case is
\begin{eqnarray}
\tilde{q}_T &=& q_T \nonumber \\
\bar{D}_T ( \omega ) &=& - \frac{1}{\kappa} |\omega | ~~~~~~~~\omega \neq
0\nonumber \\
\bar{D}_L ( \omega ) &=& - \frac{1}{\kappa} ( \omega^2 + \Delta^2 )^{1/2}
{}~~~~~~~~\omega \neq 0\nonumber \\
q_L &=& \frac{h^2}{4 \Delta} \nonumber \\
\tilde{q}_L &=& \frac{h^2}{4 \Delta} - \frac{\Delta}{\kappa \beta}
\end{eqnarray}
where $\bar{D}_T (0 ) = \bar{D}_L (0) = 0$ and
$\Delta$ and $q_T$ are determined by solution of the equations
\begin{eqnarray}
\Delta^2 &=& r + (u+v) \left( \frac{\kappa h^2}{4 \Delta} -
\frac{1}{\beta} \sum_{\omega} (
\omega^2 +
\Delta^2 )^{1/2} \right) + (M-1) v \left( \kappa q_T - \frac{1}{\beta}
\sum_{\omega}
|\omega| \right) \nonumber \\
0 &=& r + v \left( \frac{\kappa h^2}{4 \Delta} -
\frac{1}{\beta} \sum_{\omega} (
\omega^2 +
\Delta^2 )^{1/2} \right) + (u + (M-1) v) \left( \kappa q_T -
\frac{1}{\beta} \sum_{\omega}
|\omega| \right) .
\end{eqnarray}
It is not difficult to solve these equations and determine the position of
the GT boundary
by imposing the condition $q_T=0$. We give below the results of this
procedure at
$T=0$
\begin{equation}
h_{GT} = \left(\frac{2}{\pi\kappa}\right)^{1/2} \left( \frac{u (r_c  -
r)}{v} \right)^{3/4}
\log^{1/2} ( \Lambda_{\omega}^2 / \lambda_{\omega}^2 )
\label{GTres}
\end{equation}
where $\lambda_{\omega} = (r-r_c)^{1/2}$. The resulting $T=0$ phase diagram
is sketched
in Fig~\ref{GTphase}. The $T=0$ singularity in $h_{GT}$ as $r \rightarrow
r_c$ is seen to be quite
different from that in the classical model.

\subsubsection{Phases in a field coupling to the conserved total angular
momentum}
\label{conservefield}
For spin-glass models with a continuous, global symmetry, Noether's theorem
implies
that there is a conserved charge which commutes with the Hamiltonian. For
the present
quantum rotor models this charge is the total angular momentum. In this
section, we will
examine the properties of ${\cal H}_R$ in a field which couples to this
total angular momentum:
\begin{equation}
{\cal H}_R \rightarrow {\cal H}_R - \frac{1}{2} H_{\mu\nu}  \sum_{i}
\hat{L}_{i\mu\nu} .
\end{equation}
For $M=3$, $H_{\mu\nu} = \epsilon_{\mu\nu\lambda} H_{\lambda}$, for the
usual vector field
$H_{\lambda}$.
For the experimental situations in which the quantum rotor model may be
realized in
systems with short-range antiferromagnetic order~\cite{qcrit}, $H$
corresponds simply to
a uniform external magnetic field.
Obviously there is no analog of such a field for the
$M=1$ Ising case.

The properties of non-random quantum rotor models in the presence of a
field $H$ were examined
recently in Ref~\cite{conserve}, where it was found that, in high enough
dimensions, a strong $H$
always induced magnetic long-range order in a plane perpendicular to the
applied field. We will find
an analogous phenomenon here for the spin glass case.

The coupling to $H$ in the effective action can be determined by general
gauge-invariance
arguments~\cite{fishersigma,conserve}: these are equivalent to the physical
requirement that
the only effect of an applied $H$ is a uniform precession of all the
rotors. By this method,
we can deduce that the only effect of $H$ is to replace the linear, time
derivative terms in
${\cal A}$ in (\ref{landau}) by covariant derivatives:
\begin{equation}
\frac{1}{\kappa t} \int d^d x \left. \int d \tau \sum_a \left( \frac{d}{d
\tau_1 } \delta_{\mu\nu}
+ i H_{\mu\nu} \right)
\left( \frac{d}{d \tau_2 } \delta_{\mu\rho}
+ i H_{\mu\rho} \right) Q^{aa}_{\nu\rho} ( x, \tau_1 , \tau_2 )
\right|_{\tau_1 = \tau_2 = \tau}
\label{coupH}
\end{equation}
where $H_{\mu\nu} = - H_{\nu\mu}$.
Note that unlike the longitudinal field, $h$, $H$ does not couple directly
to the
replica off-diagonal components of $Q$. Related to this is the rather
straightforward
consequence of $H$ in the classical limit. This can be seen by looking at
the zero
frequency component of (\ref{coupH}) (which dominates in the classical
limit) from which we
obtain a shift  $-H_{\mu\nu} H_{\mu\rho}$ in the value of $r$
controlling spin-glass ordering in the $\nu\rho$ plane; the only effect of
$H$ is
therefore to break the ${\rm O}(M)$ invariance by inducing anisotropic
shifts in the critical
point towards ordering in different directions.

The consequences of $H$ in the quantum model are a little more interesting,
as we now see in a mean-field treatment of ${\cal A}$ in the presence of $H$.
For simplicity, we will present the analysis of the $M=3$ case only. Let us
choose $H$ to lie
along the $3$ axis, {\em i.e.\/} $H_{12} = -H_{21} = H$, with other
components zero.
We will also restrict ourselves to the paramagnetic phase in which case we
can use the
ansatz (\ref{ansatzpara}), with the modification that
$D$ has to be replaced by a general tensor $D_{\mu\nu}$ in ${\rm O}(3)$ space.
It turns out that the following parametrization of $D_{\mu\nu}$, in a
`circularly-polarized' basis
is most convenient
\begin{eqnarray}
D_{11} = D_{22} &=& \frac{1}{2} ( D_{+-} + D_{-+} ) \nonumber \\
D_{12} = -D_{21} &=& \frac{i}{2} ( D_{+-} - D_{-+} )
\end{eqnarray}
with all other components, except $D_{33}$, set equal to 0.
Inserting this result in ${\cal A}$, we find that the result (\ref{calf})
for the free energy
density is modified to
\begin{eqnarray}
\frac{{\cal F}}{n} =&& \frac{1}{t\kappa\beta} \sum_{\omega} \left[ (\omega^2 +
r) D_{33}(\omega) + ((\omega+iH)^2 + r) D_{+-}( \omega ) +
((\omega-iH)^2 + r) D_{-+} ( \omega ) \right] \nonumber \\
&&~~~~~~~- \frac{\kappa}{3t \beta} \sum_{\omega} \left[ D_{33}^3 ( \omega )
+ D_{+-}^3 ( \omega ) + D_{-+}^3 ( \omega )
\right] \nonumber \\
&&~~~~~~~ + \frac{u}{2t} \left[ \frac{1}{\beta} \sum_{\omega} D_{33}
(\omega ) \right]^2
+ \frac{u}{t} \left[ \frac{1}{\beta} \sum_{\omega} D_{+-} (\omega ) \right]
\left[ \frac{1}{\beta} \sum_{\omega} D_{-+} (\omega ) \right] \nonumber \\
&&~~~~~~~+ \frac{v}{2t} \left[ \frac{1}{\beta} \sum_{\omega} (D_{33}
(\omega ) + D_{+-} ( \omega )
+ D_{-+} ( \omega )) \right]^2 .
\end{eqnarray}
The saddle-point equations for $D_{33}$, $D_{+-}$ and $D_{-+}$ can now be
obtained and solved.
The solution is
\begin{eqnarray}
D_{33} ( \omega ) = - \frac{1}{\kappa} && (\omega^2 + \tilde{r}_1 )^{1/2}
\nonumber \\
D_{+-} ( \omega ) = - \frac{1}{\kappa} ( (\omega+iH)^2 + \tilde{r}_2
)^{1/2} ~~~~~~~~&&~~~~~~~~~
D_{-+} ( \omega ) = - \frac{1}{\kappa} ( (\omega-iH)^2 + \tilde{r}_2 )^{1/2}
\label{resdpm}
\end{eqnarray}
where $\tilde{r}_1$ and $\tilde{r}_2$ are to be determined from the
solutions to the equations
\begin{eqnarray}
\tilde{r}_1  &=& r - \frac{u+v}{\beta} \sum_{\omega} (\omega^2 +
\tilde{r}_1 )^{1/2}
- \frac{2v}{\beta} \sum_{\omega} ((\omega+ i H)^2 + \tilde{r}_2 )^{1/2}
\nonumber \\
\tilde{r}_2  &=& r - \frac{v}{\beta} \sum_{\omega} (\omega^2 + \tilde{r}_1
)^{1/2}
- \frac{u+2v}{\beta} \sum_{\omega} ((\omega+ i H)^2 + \tilde{r}_2 )^{1/2}
\label{tilder1r2}
\end{eqnarray}
which generalize (\ref{tilder}). The only effect of $H$ has been to shift
frequency
$\omega$ by $iH$ (corresponding to transforming to a rotating reference
frame in
Euclidean time) for fluctuations in the $12$ plane.

It is useful to have a better understanding of the $H$
dependent frequency summations above. The following identity, obtained from the
Poisson
summation formula, is helpful:
\begin{eqnarray}
\frac{1}{\beta} \sum_{\omega} ( (\omega + i H)^2 + \Delta^2 )^{1/2}
= &&\int_{-\Lambda_{\omega}}^{\Lambda_{\omega}} \frac{d \omega}{2 \pi}
(\omega^2 + \Delta^2)^{1/2} \nonumber \\
&&~~~ - \frac{1}{\pi} \int_{\Delta}^{\infty} d \Omega
\sqrt{\Omega^2 - \Delta^2} \left( \frac{1}{e^{\beta (\Omega-H)} - 1}
+ \frac{1}{e^{\beta(\Omega + H)} - 1}
\right)
\label{freqsumh}
\end{eqnarray}
valid for $\Delta > H$. The physical meaning of the second term in the
above equation should also be
clear: this is the contribution of the thermally-excited, circularly
polarized states
with zero-field energy $\Omega$, and density of states $\sqrt{\Omega^2 -
\Delta^2}$; these states
have been split
into states with energy $\Omega \pm H$ in the presence of a field.
Notice that at $T=0$, (provided $H< \Delta$), the second term in
(\ref{freqsumh}) vanishes,
and the
entire contribution comes from the first term which is independent $H$.
This allows us to immediately
obtain an important $T=0$ solution to the Eqns (\ref{tilder1r2}):
\begin{equation}
\tilde{r}_1 = \tilde{r}_2 = \tilde{r}~~~\mbox{at $T=0$, provided $H <
\sqrt{\tilde{r}}$}
\end{equation}
where $\tilde{r}$ is the solution of the $H=0$ equation (\ref{tilder}).
This solution could have been anticipated a priori---at $H=0$ we had
an ${\rm O}(3)$ singlet paramagnet ground
state with a gap $\sqrt{\tilde{r}}$, and the energy and wavefunction of
such a state will
not be modified by a uniform external field. However, when $H >
\sqrt{\tilde{r}}$, the energy of an
excited spin
1 state will go below that of the ground state.
By analogy with the non-random case~\cite{conserve}, we then expect
condensation into the
spin-1 state, and appearance of spin-glass order the plane perpendicular to
the field -
$q^{ab}_{11} = q^{ab}_{22} \neq 0$ for $a\neq b$.
The phase boundary to the appearance of this spin-glass ordering is
$H=\sqrt{\tilde{r}}$,
which at $T=0$ gives us from (\ref{logr})
\begin{equation}
H = \left(
\frac{4 \pi}{(u + Mv)} \frac{ (r-r_c )}{ \log(\Lambda_{\omega}^2 /
(r-r_c))} \right)^{1/2}.
\label{conserveres}
\end{equation}
This is sketched in the $T=0$, finite $H$ phase diagram of
Fig~\ref{conservefig}.
This finite $H$ spin glass/paramagnet quantum transition is in a different
universality class
from the $H=0$ transition. Some of its critical properties can be deduced
from the results above,
using methods very similar to those described in this paper for the $H=0$
transition. We will not
go into any details on this issue in this paper, apart from noting that
examining
the low $\omega$ behavior of (\ref{resdpm}) tells us immediately that the
finite $H$
quantum transition has the mean-field critical exponents $z=4$ and $\nu=1/4$.

A second phase boundary
in Fig~\ref{conservefig} at $H=H_3 ( r)$ marks
the appearance of
of spin-glass order in $q^{ab}_{33}$. Ordering in $q_{33}$ is clearly
present for $r<r_c$ at $H=0$,
and should therefore also be present for small enough $H < H_3 (r)$.
Determining $H_3$ requires
computations in the spin glass phase, with replica off-diagonal components
of $Q$ non-zero.
As the computation is rather similar to those already carried out, we will
not present any details
and simply state the final result. We find
\begin{equation}
H_3 = \left( \frac{u}{2 v} ( r_c - r) \right)^{1/2} .
\label{H3res}
\end{equation}

We close our discussion of $H$ by examining the special point $r=r_c$.
At finite $H$ and $T$, we expect a transition to a spin-glass phase when
$H/T$ is greater than a
universal number~\cite{conserve} (which can again be determined from
(\ref{tilder1r2})).
For $H \ll T$ however, the paramagnetic phase is stable and we can perform
an expansion in
powers of $H$ to obtain the finite $T$ spin susceptibility of the
quantum-critical point.
An expansion of the free energy density in the paramagnetic phase in powers
of $H$ yields
\begin{equation}
\frac{{\cal F}(H) - {\cal F}(H=0)}{n} = -\frac{\beta H^2}{\pi \kappa^2 t}
\int_{\sqrt{\tilde{r}}}^{\infty} d \Omega \frac{ \Omega (\Omega^2 -
\tilde{r})^{1/2}}{\sinh^2
(\beta \Omega /2)}
\end{equation}
where $\tilde{r}$ is determined by the solution of $H=0$ equation
(\ref{tilder}).
Evaluating the integral at $r=r_c$ and at low $T$ (regime II of
Fig~\ref{phasediag}) we find
\begin{equation}
\frac{{\cal F}(H) - {\cal F}(H=0)}{n} = -\frac{4 \pi H^2 T^2 }{3 \kappa^2 t} .
\label{freeenh}
\end{equation}
The susceptibility, $\chi_H$ associated with $H$ therefore behaves as $\sim
T^2$ at low
temperatures. We may combine this result with $T$ dependence of the
specific heat in
(\ref{freeent}) to obtain the dimensionless, universal Wilson
ratio~\cite{conserve},
$W$,
characterizing the `quantum-critical' regime II of Fig~\ref{phasediag}:
\begin{equation}
W = \frac{T \chi_H}{C_V} = \frac{5}{6 \pi^2} .
\end{equation}

\section{Renormalization group analysis}
\label{renormgroup}
This section will attempt to go beyond the mean-field results of
Section~\ref{landautheory} by subjecting the finite dimensional theory to a
Wilson style
renormalization group (RG) which integrates out degrees of freedom at high
energies and
large momenta.

We will study the properties of ${\cal A}$ under the rescaling transformation
\begin{equation}
x^{\prime} = x/b~~~~~~~~\tau^{\prime} = \tau / b^{z}
\end{equation}
where $z$ is the dynamic critical exponent.
The coupling $t$ will play a special role in the following, and we find it
convenient to
give it its own independent rescaling transformation
\begin{equation}
t^{\prime} = t b^{-\theta} .
\label{transt}
\end{equation}
The exponent $-\theta$ will be the scaling dimension of $t$ near a fixed
point with $t=0$;
all fixed points found below in fact will have $t=0$, and $\theta>0$,
making $t$ a (dangerously)
``irrelevant'' coupling.
As is conventional, we define the anomalous
dimension $\eta$ such that $\eta=0$ in the approximation in the previous
section; in RG this
means that the coefficient of the $(\nabla Q)^2$ term is not rescaled at
tree level. This
imposes the field rescaling
\begin{equation}
Q^{\prime} = Q b^{(d-\theta+2z-2+\eta)/2} .
\label{transQ}
\end{equation}
The exponents $z$ and $\theta$ will be fixed by demanding that the
transformations
of the $1/\kappa t$ linear term with time derivatives, and the $1/t^2$
quadratic
coupling, are consistent with (\ref{transt}) and the RG equation for the cubic
$\kappa /t$ coupling.

It is now a simple matter to determine the tree level rescalings of the
remaining couplings
(we determine the rescaling of $\kappa$ from the cubic term):
\begin{eqnarray}
r^{\prime} = r b^{2z}~~~~~~~~&&~~~~~~~~~\kappa^{\prime} = \kappa
b^{(6+\theta-d-3\eta)/2}\nonumber \\
u^{\prime} = u b^{2-z-\eta}~~~~~~~~&&~~~~~~~~~v^{\prime} = v b^{2-z-\eta} .
\label{rescalings}
\end{eqnarray}
The full action remains invariant at tree level under these transformations
if we choose
\begin{equation}
z = 2~~~~~~~~~~~~~~\eta=0~~~~~~~~~~~~~~~\theta=2 .
\end{equation}
We see then from (\ref{rescalings}) that $u$ and $v$ are always marginal, while
$\kappa$ becomes relevant below $d=8$. As $\theta$ is positive, $t$ is
irrelevant and
flows towards $t=0$.

Before we can reach any conclusions on the meaning of these scaling
dimensions, we need to
understand the role of the coupling $t$ in the loop corrections. Recall
that we determined in
Section (\ref{paracrit}) that the propagator, $G$, came with a prefactor
$t$, the cubic interaction
was $\kappa/t$, and the quadratic couplings are $u/t$, $v/t$ and $1/t^2$
(Fig~\ref{vertices}). The
presence of inverse powers of $t$ leaves open the possibility that the loop
corrections acquire even
higher powers of $1/t$. In fact, it is not difficult to check that this
does {\em not\/} happen.
The argument is similar to that in the classical problem of a Ising
ferromagnet in a
random field~\cite{geoff,dsf}: there are always enough $t$ factors
in the numerator from $G$ to cancel those from the couplings. In
particular, there can be only
one insertion of the $1/t^2$ quadratic coupling in any propagator, as two
of them will involve
a vanishing trace over replica indices.
This discussion can be summarized in terms of a simple result:
all loop corrections to any coupling or Green's function
are no more singular as $t\rightarrow 0$ than the results at tree level.
The coefficient of
this leading singular power of $t$ will however contain an infinite number
terms with positive
powers of the couplings $u$, $v$, and $\kappa$. These three couplings,
therefore, do indeed measure the
strength of the loop corrections.
Above $d=8$, the cubic $\kappa$ coupling is irrelevant, and we expect
that the mean-field and tree-level results of Section~\ref{meanfieldtheory}
are related to
those of a Gaussian fixed point.

Now we turn to determining loop corrections to the renormalization group
equations.
We will obtain the one-loop flow equations by integrating out degrees of
freedom
with frequencies $\omega$ such that $\Lambda_{\omega} /b^z < \omega <
\Lambda_{\omega}$
{\em or\/} momenta $k$ such that $\Lambda_k /b < k < \Lambda_k$, where
$\Lambda_k$ is
a cutoff in momentum space.
We determine the saddle-point of the action with respect to variations in
$Q$ fields in this range of momenta and frequencies {\em only\/}, and obtain
fluctuation corrections to one-loop.
 This is followed by the rescaling transformation discussed
above. It is also useful to keep track of the free energy density ${\cal
F}$ in this procedure:
we introduce an overall constant to ${\cal A}$
\begin{equation}
{\cal A} \rightarrow {\cal A} +  \int d^d x d \tau f
\end{equation}
and keep track of the flow equation of $f$.
We will obtain the differential form of the scaling equations with $b =
e^{\ell}$.

The diagrams which contribute at one loop order are shown in Figs~\ref{loops}
and~\ref{kappa}.  There are two types of diagrams at this order.
The first have a loop with
an internal momentum (Figs~\ref{loops}a,c and~\ref{kappa})
which must be integrated over,
with all frequencies pinned to the external ones: these are associated with
a momentum space
factor of $K_k \ell
\equiv S_d
\ell / (2 \pi)^d$ where $S_d$ is the surface-area of a sphere in $d$
dimensions.  The
second (Fig~\ref{loops}b) are really tree-level diagrams but have a
frequency integration
because of quantum-mechanical interactions: these have a frequency space
factor of
$K_{\omega} \ell \equiv 2 z \ell / (2 \cdot 2 \pi)$ (the factor of $2$ in
the numerator comes from the
frequency shells at $\pm \Lambda_{\omega}$, and the $2$ in the denominator
is from the propagator which
at zero momentum is $1/2\sqrt{\omega^2 + r}$).
It is convenient to absorb these factors into the coupling constants by the
transformations $K_{\omega} u \rightarrow u$, $K_{\omega} v \rightarrow v$,
$K_k \kappa^2 \rightarrow \kappa^2 $, $r/ \Lambda_{\omega}^2 \rightarrow r$ and
$f \rightarrow K_{\omega} \Lambda_{\omega}^4 f$.
Evaluating the diagrams by completely standard methods we obtain the
following flow equations (dropping innocuous factors of $\Lambda_k^{d-8}$ in
the terms with a momentum integration):
\begin{eqnarray}
z &=& 2 + 4 \kappa^2 \nonumber \\
\eta  &=&  2 \kappa^2 \nonumber \\
\theta &=& 2- \frac{2(u+Mv)}{\sqrt{1+r}} \nonumber \\
\frac{dt}{d\ell} &=& - \theta t \nonumber \\
\frac{dr}{d\ell} &=& 2 z r - 2(u + Mv) \sqrt{1 + r} -
\frac{\Lambda_k^4}{\Lambda_{\omega}}
\kappa^2
\nonumber \\
\frac{d\kappa }{d \ell} &=& \frac{8-d}{2} \kappa + 9 \kappa^3 - \frac{ (u + Mv)
\kappa}{\sqrt{1+r}}
\nonumber \\
\frac{du}{d \ell} &=& - 2  u \kappa^2 - \frac{u^2}{\sqrt{1+r}} \nonumber \\
\frac{dv}{d \ell} &=& -2 v \kappa^2 - \frac{2 u v + M v^2 }{\sqrt{1+r}}
\nonumber \\
\frac{df}{d\ell} &=& (d + z) f - \frac{4}{3 t\kappa^2} (1 + r )^{3/2} .
\label{rgeqns}
\end{eqnarray}
[As an aside, we note here that the dependence of the flow equations (and
therefore the critical
exponents) on $M$ is illusory. The couplings $u$ and $v$ enter into the
equations for
the other couplings only in the combination $U \equiv u + M v$. The
equation for $U$ is
in fact
$M$ independent:
\begin{equation}
\frac{d U } {d \ell} = - 2 U \kappa^2 - \frac{U^2}{\sqrt{1 + r}}.
\label{rguueqn}
\end{equation}
We may trade the couplings $u,v$ for $U,u$, and the equations for the latter
are $M$
independent. This independence on $M$ is similar to that found in the
infinite-range
model~\cite{letter}.]

We begin by determining all the fixed points of these flow equations.
The parameter $r$ is obviously associated with the thermal operator, and
has relevant flows
from all the fixed points we find. The value of $r$ at these fixed points is
of order $\epsilon = 8-d$, and to leading order in $\epsilon$ this value of
$r$ does not
feed into the positions of the other couplings. We can therefore set $r=0$,
and focus on the
other couplings. We find 5 fixed points; their positions to leading order
in epsilon are
listed below, along with the three eigenvalues controlling flows near the
fixed point in
$\kappa,u,v$ space (there is, of course, also the omnipresent relevant
eigenvalue associated
with $r$). All the fixed points have $t=0$ and the eigenvalue controlling
flow in the
$t$ direction is $-\theta$.
\begin{eqnarray}
(A)~~~&& \kappa=0, u=0, v=0~~~~~~~~~~~~~~~~~~~~~~(\epsilon/2,0,0) \nonumber \\
(B)~~~&& \kappa^2 =-\epsilon/22,  u=\epsilon/11 ,
v=0~~~~~~~~~~~(-(5\pm\sqrt{14})\epsilon/11, -\epsilon/11) \nonumber \\
(C)~~~&& \kappa^2 =-\epsilon/22, u=0,
v=\epsilon/11M~~~~~~~~(-(5\pm\sqrt{14})\epsilon/11,
\epsilon/11) \nonumber \\
(D)~~~&& \kappa^2=-\epsilon/18, u=0,
v=0~~~~~~~~~~~~~~~~(-\epsilon,\epsilon/9,\epsilon/9)\nonumber
\\ (E)~~~&& \kappa^2=-\epsilon/18, u=\epsilon/9, v=-\epsilon/9M~~~~~
(-\epsilon,\epsilon/9,-\epsilon/9) .
\label{rgfixedpoints}
\end{eqnarray}
The properties and stabilities of these fixed points depend strongly on the
sign of
$\epsilon$, and we will consider $\epsilon < 0$, $\epsilon =0$ and
$\epsilon > 0$ separately.

\subsection{$d > 8$}
While all five fixed points have real values of $\kappa$,
the only stable fixed point is the Gaussian fixed point $A$.
Related to the two zero eigenvalues, the couplings $u$ and $v$ are both
marginal.
The renormalization group flows are shown in the $\kappa, U=u+Mv$ plane in
Fig~\ref{rgflowdgt8}. Observe that the basin of attraction of $A$ is
limited to a portion of the
$U > 0$ quadrant (the basin of attraction is actually also restricted to $u>0$,
$v>0$). For initial values of the couplings outside this basin, we have
runaway flows to
strong couplings and are therefore unable to make any firm predictions. The
rest of the
discussion in this subsection will therefore be limited to initial values
within the
basin attraction of $A$. Some speculations on the properties of the
remainder of the
phase space will be made in Sec~\ref{scalinghypoth}.

The properties of $A$ are obviously related to those of the mean-field theory
of
Section~\ref{meanfieldtheory}. The logarithmic corrections found there can
be attributed to the
couplings $u,v$ which are marginal near $A$ for {\em all\/} $d>8$; this
will be shown below. First we focus
on the scaling properties, modulo logarithms. The `thermal' coupling $r$
has eigenvalue $4$, leading to the
critical exponent  \begin{equation}
\nu=\frac{1}{4}.
\label{resnu}
\end{equation}
We also know from the fixed point values of $\kappa$, $u$, and $v$, and
(\ref{rgeqns}) that
\begin{equation}
z=2~~~~~~~~~~~~~~~\eta=0~~~~~~~~~~~~~\theta=2
\label{resz}
\end{equation}
as found in the mean-field theory of Sec~\ref{paracrit}.
There are two irrelevant couplings, $t$ and $\kappa$, associated with
eigenvalues
$-\theta$, and $-\theta_{\kappa}/2$ respectively, with
\begin{equation}
\theta_{\kappa} = d-8.
\label{thetakappa}
\end{equation}
Now we turn to the Green's functions of Section (\ref{observables}).
Their scaling dimensions will be given by those of $Q$ in (\ref{transQ}) and
any
dependence they have on the irrelevant $t$, $\kappa$ couplings.
{}From the results of
Section (\ref{paracrit}) we see that  $D \sim 1/\kappa$ (Eqn (\ref{resd})),
$G \sim t$ (Eqn (\ref{propagator})), $G^c \sim t$ (Eqn (\ref{resgc})), and
$G^{d} \sim t^0 $ (Eqn (\ref{resgd1})). Combining
this information with the scaling dimension of $Q$ (Eqn (\ref{transQ})), we
obtain the following
results for the scaling dimensions of the Green's functions {\em after\/} their
arguments
have been Fourier transformed to $k$ and $\omega$:
\begin{eqnarray}
\mbox{dim} ( D) &=& (d-\theta-\theta_{\kappa}-2 +\eta)/2 \nonumber \\
\mbox{dim} ( G) &=& -2 + \eta  \nonumber \\
\mbox{dim} (G^{d}) &=& -2 -\theta + \eta \nonumber \\
\mbox{dim} (G^c ) &=& -2 -z + \eta .
\end{eqnarray}
Note that these results are consistent with those in
(\ref{propagator})--(\ref{defbareta})
and that
\begin{equation}
\bar{\eta}
= \eta + 2 - \theta.
\end{equation}

The free energy also has a singular dependence upon $\kappa$ and $t$:
${\cal F} \sim
1/t\kappa^2$ (see Eqn (\ref{resfreepara})). Related to this is the singular
$1/t\kappa^2$ term
in the renormalization group flow equation for $f$ in (\ref{rgeqns}). From
this flow equation
it is easily seen that
\begin{equation}
\mbox{dim} ({\cal F}) = d + z - \theta - \theta_{\kappa} .
\end{equation}
Thus hyperscaling is violated; notice that
$\theta$ does not vanish as $d$ approaches 8 from above, so violation of
hyperscaling continues
to occur even in the upper critical dimension. This result, with the
exponent values in (\ref{resnu}),
(\ref{resz}) and (\ref{thetakappa}), predicts a $d$-independent singular
$(r-r_c)^2 $ term in
${\cal F}$, consistent with (\ref{resfreepara}).

Next, we examine the form of the logarithmic corrections to scaling which
are present in all
$d>8$. For $d-8$ of order unity, the irrelevant couplings $\kappa$,$t$
decrease rapidly, and can
therefore be set equal to $0$ in the renormalization group equations
(\ref{rgeqns}), (\ref{rguueqn}) (except
in the equation for $f$, where they appear in the denominator). For small
$r$, we can then integrate
(\ref{rguueqn}) and find
\begin{equation}
U ( \ell ) \sim \frac{1}{\ell}
\end{equation}
for large $\ell$. We have to now insert this into the equation for $r$ in
(\ref{rgeqns}), and
integrate up to the scale $\ell = \ell^{\ast}$ at which $r(\ell^{\ast})
\approx 1$.
Doing this to leading logarithmic accuracy we find that
\begin{equation}
\ell^{\ast} e^{- 2 z \ell^{\ast}} \approx r(\ell = 0) - r_c .
\end{equation}
The typical frequency scale near the critical point is $e^{-z \ell^{\ast}}$
which from the above is
of order $[(r-r_c)/\log(1/(r-r_c))]^{1/2}$; this is identical to that
obtained in (\ref{logr}) in the
mean-field theory of Sec~\ref{paracrit}. A similar analysis can be
undertaken for the free energy.
%establishing our claim that the logarithmic corrections in
Sec~\ref{paracrit} are due to the marginal
%couplings $u$ and $v$.

Finally, recall that in the mean-field results within the spin-glass phase
in Secs~\ref{spinglassphase}
and~\ref{longitudinalfield}, all replica symmetry breaking effects were
associated with the formally
irrelevant coupling $y_1$ in (\ref{ycouplings}). The scaling dimension of
the $y_1$ perturbation is
$ - (d + 2 z - \theta -4 + 2 \eta)$; knowing this and the scaling
dimensions above, the mean-field critical
properties of the replica-symmetry breaking effects can be given a standard
scaling interpretation.

\subsection{$d=8$}
The flows are shown in Fig~\ref{rgflowdeq8}. The only stable fixed point
remains $A$ and
its basin of attraction is the region $\kappa^2 \leq U/20$. Now the flow of
$\kappa$ into this
fixed point is also marginal (in addition to $u,v$) and so there will be
new logarithmic
corrections to scaling. As before, we know little about the system outside
the basin of $A$.

\subsection{$d<8$}
Now $A$ is the only fixed point with $\kappa$ real, but is unstable.
The flows are shown in Fig~\ref{rgflowdlt8}.
Thus over the entire,
physical region we have flows to strong coupling. There is no simple expansion
of the critical properties in powers of $\epsilon$. No definite results can
thus be
obtained, and we resort to some speculations using phenomenological scaling
ideas in the
next section.

\section{Scaling Hypotheses}
\label{scalinghypoth}

In this section we discuss the scaling properties that may be expected at
the transition in our
models; as far as possible this will be done without reference to the
Landau theory or to replicas.
We begin by collecting the definitions of exponents and the scaling hypotheses
that have been
mentioned at various points earlier in the paper.

Because of the importance of the dangerously irrelevant (DI) variable $t$
in the Landau theory and
to gain greater generality, we will assume that such a variable, also
denoted $t$ with scaling
dimension $-\theta < 0$, exists in the scaling theory. We assume that there
is only this one DI
variable; the coupling $\kappa$ that is also DI in Landau theory for $d >
8$, will not be for $d <
8$, as in ordinary critical phenomena. The extension to include more DI
variables is
straightforward, and of course if none are present, one can simply set
$\theta =0$. Moreover, it
will be assumed that $t$ appears in a similar way as in the Landau theory,
which contains a $1/t^2$
term as well as $1/t$ terms. This means that in each realization of
disorder, $t$ plays a role
similar to $\hbar$ in quantum field theory (or $T$ in classical statistical
mechanics): when it is
small, certain types of disorder-induced fluctuations (those directly
responsible for determining the local
position of the critical point) dominate. (The analogy is not exact because
of the
internal frequency integrals that can occur even in tree diagrams in $Q$
within Landau theory.)
%Thus
%the assumption that $t$ is dangerously irrelevant means that fluctuations
due to quenched disorder
%dominate over (some of) the quantum fluctuations near the critical point.

Defining $G^d$ as in (\ref{defgd1}), we can then define the exponent
$\bar{\eta}$ by
\begin{equation}
G^d (x, \tau, \tau) \sim x^{-(d+2z-4+\bar{\eta})}
\end{equation}
for fixed $\tau/x^z$ at criticality; there is no dependence on $t$ on the
right hand side. Thus the
dimension of $Q \sim S S$ when the spins are separated in time is $(d + 2z
- 4 + \bar{\eta} )/2$.
Therefore also (recall (\ref{defD}))
\begin{equation}
D \sim \tau^{-(d+2z-4+\bar{\eta})/2z} .
\end{equation}
The spin glass correlator, $G$, (recall (\ref{defgd2})) vanishes if $t=0$,
so is proportional to $t$
(as $t \rightarrow 0$) and thus behaves as
\begin{equation}
G (x, 0, 0) \sim x^{-(d + 2z - 4 + \bar{\eta} + \theta)} .
\end{equation}
Comparing with the definition of $\eta$, $G \sim x^{-(d + 2z - 2 + \eta)}$
yields
\begin{equation}
\theta = 2 + \eta - \bar{\eta}.
\label{thetaeta}
\end{equation}
{}From these we may obtain other scaling relations, {\em e.g.}\ the
spin-glass susceptibility
$\chi_{SG} \sim (r - r_c)^{-\gamma}$ with
\begin{equation}
\gamma = (2 - \eta)\nu,
\end{equation}
and the order parameter $q = \left[ \left\langle S \right\rangle^2 \right]
\sim (r_c - r)^{\beta}$
with
\begin{equation}
\beta = (d + 2z - \theta -2 + \eta) \frac{\nu}{2}.
\end{equation}
In general, the usual scaling relations are obeyed, except that
hyperscaling involves $d+2z -\theta$
in place of $d$ for classical critical phenomena whenever the bilocal field
is involved (hence the
$2z$) and the $\theta$ is due to the DI variable $t$. On the other hand, a
field that is local in
time behaves normally (on including $z$ and $\theta$). Thus the free energy
density scales
as $(r-r_c)^{(d+z-\theta)\nu}$ as $r \rightarrow r_c$, and the specific
heat at finite temperature $T
\rightarrow 0$ at $r=r_c$ behaves as $T^{(d-\theta)/z}$; the values $z=2$,
$\theta=2$ yield
$T^3$ at $d=d_u = 8$ in the Gaussian theory, as was obtained directly in
the infinite-range
model~\cite{letter} and from the mean field theory in (\ref{freeent}). The
response of the system to
an external field coupling to the conserved angular momentum scales as
$T^{(d-\theta-z)/z}$~\cite{conserve} in the quantum-critical region II of
Fig~\ref{phasediag}---at
$d=8$ this is $T^2$, in agreement with (\ref{freeenh}).

A  local variable, which has not been introduced so far, but is important
for the present scaling
considerations, is the ``thermal'' operator
$\psi (x, \tau)$. This variable couples to the control parameter $r$ that
tunes the system across
the quantum transition:
\begin{equation}
\psi (x, \tau) = S_{\mu} ( x, \tau) S_{\mu} ( x, \tau)
\end{equation}
in a single replica of the system. For the disconnected correlator of $\psi$ we
may define at
criticality
\begin{equation}
\left[\left\langle \psi (x , \tau_1 ) \right\rangle \left\langle
\psi (0, \tau_2 ) \right\rangle \right] \sim x^{-(d-4+\bar{\eta}_{\psi})}
\end{equation}
(so that the Fourier transform $\sim k^{-4 + \bar{\eta}_{\psi}}$; note the
correlator is
independent of
$\tau_1$, $\tau_2$) so that the scaling dimension of $\psi$ is $(d-4+
\bar{\eta}_{\psi})/2$. The
connected correlation function then goes as $\sim t$ and so
\begin{eqnarray}
\left[\left\langle \psi (x , 0)
\psi (0, 0) \right\rangle \right] -
\left[\left\langle \psi (x , 0) \right\rangle \left\langle
\psi (0, 0) \right\rangle \right] &\sim&
x^{-(d-4+\theta+\bar{\eta}_{\psi})} \nonumber \\
&\equiv& x^{-(d+z-2+ \eta_{\psi})}
\end{eqnarray}
(from $k^{-2+\eta_{\psi}}$ in Fourier space) and hence
\begin{equation}
\theta = 2 + z + \eta_{\psi} - \bar{\eta}_{\psi} .
\label{thetapsi}
\end{equation}
As usual the dimension of $\psi$ determines that of $r-r_c$, and hence the
value of $\nu$, through
modified hyperscaling, that is
\begin{eqnarray}
\frac{1}{\nu} &=& d+ z - \theta - \frac{1}{2} ( d - 4 + \bar{\eta}_{\psi}
)\nonumber\\
&=& \frac{1}{2} ( d + z - \theta + 2 - \eta_{\psi} ) .
\label{nupsi}
\end{eqnarray}

An interesting rigorous inequality was proved by Schwartz and
Soffer~\cite{schwartz} for the
exponents $\eta_{\psi}$, $\bar{\eta}_{\psi}$ in any system where a local
field $\psi$ couples to
Gaussian disorder, as is the case for our $\psi$. Extending their proof to
the case where the
disorder is time independent, we obtain
\begin{equation}
\bar{\eta}_{\psi} \leq 2 \eta_{\psi}
\label{etapsilteta}
\end{equation}
and hence from (\ref{thetapsi}) $ \theta \geq 2 + z - \eta_{\psi}$. Using
(\ref{nupsi}) this implies
\begin{equation}
\nu \geq \frac{2}{d} .
\label{nugeq2d}
\end{equation}
This inequality was proved by Chayes {\em et. al.\/}~\cite{harris}. The
present approach appears
easier but rests upon the use of a scaling relation to obtain $\nu$.

For the classical random field Ising model, it has been claimed that
$\bar{\eta}_{\psi} =
2 \eta_{\psi}$ is satisfied as an equality~\cite{schwartz2}.
This would imply that the  correlation
length at $T=T_c$ due to a uniform field $h$ would go as $\xi \sim
h^{-2/d}$. However we find the
proof unconvincing, though series results do seem to show that the equality
is accurately obeyed in
$d=3,4,5$ in that problem.

Since $\psi = S S$ it is tempting to equate the scaling dimensions of $Q$
and $\psi$
($d+2z-4+\bar{\eta} \stackrel{?}{=} d - 4 + \bar{\eta}_{\psi}$) and obtain
using (\ref{thetaeta}) and
(\ref{thetapsi})
\begin{equation}
\eta_{\psi} \stackrel{?}{=} z + \eta .
\end{equation}
However $\psi$ involves bringing spins $S$ to the same time as well as
position (and summing over
spin indices) and so may have different renormalizations than $Q$. Thus we
do {\em not\/} expect
this relationship to hold.

Finally, let us consider the connected correlation function, $G^c$, defined in
(Eqn (\ref{defgc})). Similar scaling ideas suggest the following form at
criticality
(taking the
$M=1$ case for simplicity)
\begin{equation}
G^c (x, \tau, 0, \tau) \sim \left( x^{2z} + \tau^2 \right)^{-(d+2z-2+\eta)/2z}
\Gamma^c ( \tau /x^z )
\label{gcgamma}
\end{equation}
with $\Gamma^c (y)$ a universal function. Taking either of the limits
$\tau\rightarrow 0$,
or $x \rightarrow 0$ has the consequence of bringing 2 of the 4 $S$ fields
in $G^c$ to the same
spacetime point, so that $G^c$ reduces to a correlator of $\psi$. Knowing
the scaling dimension
of $\psi$, this procedure fixes the asymptotic limits of $\Gamma^c$:
\begin{equation}
\Gamma^c (y) \sim \left\{
\begin{array}{ll}
y^{(\eta_{\psi} - z - \eta )/z} & \mbox{as $y \rightarrow 0$} \\
y^{(\eta + z - \eta_{\psi} )/z} & \mbox{as $y \rightarrow \infty$} .
\end{array}
\right.
\label{gcetapsi}
\end{equation}
Similar statements hold for the general $G^{c} (x, \tau_1 , \tau_2 , \tau_3
)$, and
for $G^d$ with a different scaling function $\Gamma^d$.

We now compare the above relations to Monte Carlo results for $M=1$ in
$d=2$~\cite{rieger}
and $d=3$~\cite{guo}. Their results are $\nu^{-1} \approx 1.3$, $z \approx
1.3$ ($d=3$)
and $\nu^{-1} \approx 1$, $z \approx 1.5$ ($d=2$). They examined scaling of
several
susceptibilities, most of which are related to $G$ and hence involve $\eta$
in straightforward
ways; however their definition of $\eta$ is what we denote $\eta^{\prime} =
\eta + z$.
In our notation their results are $\eta \approx -0.2 $ ($d=3$), $\eta
\approx -1.0 $
($d=2$). (We have corrected an arithmetical error in the paper Guo {\em et.
al.\/}~\cite{guo}: $\eta^{\prime} = 1.1$, not $0.9$.) They also examined
$\chi_{nl}$
which is   related to $G^c$ (Eqn
(\ref{defchinl})) and find its scaling dimension is consistent with
the assumption that the scaling forms like (\ref{gcgamma}) can be
integrated so that
$\eta_{\psi}$ drops out and $\chi_{nl} \sim L^{z+2-\eta}$, and using
the same $\eta$ as the other susceptibilities. Note that the negative value
of $\eta$ is quite
reasonable in a disordered system~\cite{qcrit}.

The numerical results so far give no test of hyperscaling, but Guo {\em et.
al.\/}~\cite{guo}
also studied $D$, obtaining $D(\tau) \sim \tau^{-1.3}$. Using the scaling
relations and values of
exponents in $d=3$ we obtain $\theta \approx 0.0$. This may mean that
conventional hyperscaling is
obeyed, though because of uncertainties in exponents, a small positive
$\theta$ cannot be ruled out.
Clearly more work on this point, and tests of other scaling relations,
would be welcome.
(If $\theta=0$ then $\eta_{\psi} = z + \eta$ is ruled out.) It is
interesting that in both
$d=2$ and $3$ , $\nu^{-1} \approx d/2$ and this also holds {\em exactly\/}
in the $d=1$
model~\cite{daniel}---this raises the question whether the inequalities
(\ref{etapsilteta}),
(\ref{nugeq2d}) are saturated; at present we have no argument why this
might be so.

\section{Conclusions}
In this section we summarize our main results and discuss their
relationship to other open problems.
A comparison of our results with some recent work has already been presented in
Sec~\ref{scalinghypoth}.

We have studied  models of quantum rotors or Ising spins in a transverse field
with random, short-range, frustrating exchange interactions. We examined
properties of these systems in the vicinity of a zero temperature quantum
transition between a
spin-glass and a quantum paramagnet phase. We characterized this transition
by an order
parameter field $Q_{\mu\nu}^{ab} (x, \tau_1 , \tau_2)$ which is a matrix in
spin components ($\mu$,
$\nu$) and replica indices ($a,b $), and is {\em bilocal\/} in Matsubara
time ($\tau_1$, $\tau_2$).
The expectation value of $Q$ is the on-site two-spin correlation function
which becomes long-range
in time at the onset of spin-glass order. We then introduced a Landau
effective action functional
for $Q$: the functional was written down as the most general one consistent
with a set of symmetry
criteria and the usual Landau theory requirements of locality in space and
time. The functional involves {\em
both\/} the replica diagonal and off-diagonal components of $Q$---contrast
this with classical spin
glass~\cite{youngbinder,book} for which the Landau theory involves only the
replica off-diagonal components of
an order parameter matrix field which is also independent of time.

A mean-field functional minimization of the Landau action yielded a great
deal of structure.
For parameters favoring large quantum fluctuations, we found a paramagnetic
phase whose
properties (as well as those of the quantum-critical point at which the
paramagnetic phase
terminates) were identical to those obtained in an earlier exact solution
of a model with
infinite-range interactions~\cite{miller,letter}. In systems with weaker
quantum fluctuations,
we found
a replica-symmetric spin glass ground state, with replica symmetry breaking
appearing at any non-zero temperature. We were able to study systematically
the behavior of replica
symmetry breaking at small
$T$, in contrast to the classical Sherrington-Kirkpatrick model where the
order parameter is of
order unity as $T\rightarrow 0$, and no Landau expansion exists (see
however Ref.~\cite{temesvari}).
In the present situation, we used the proximity to a quantum-critical point
to our advantage, and
developed a Landau expansion valid even at $T=0$. The response of the
spin-glass and paramagnetic
phases to a variety of external fields was also studied.

Next we examined consequences of fluctuations about mean-field for the
critical properties of the
quantum transition. We identified $d=8$ as the upper critical dimension.
 Above
$d=8$, and with certain restrictions on the values of the Landau couplings,
we found that the
transition was controlled by a Gaussian fixed point with mean-field
critical exponents. For
couplings not attracted by the Gaussian fixed point above $d=8$, and for
all physical couplings
below $d=8$, we found  runaway renormalization
group flows to strong coupling. An important feature of the renormalization
group analysis was
the appearance of a dangerously irrelevant coupling (even below $d=8$),
which played a role similar,
though {\em not\/} identical, to Planck's constant $\hbar$. As a result
certain disconnected
correlation functions measuring fluctuations due to quenched disorder were
found to be more singular
than the corresponding connected correlations which contain
quantum-mechanical fluctuations. The
structure implied by this dangerously irrelevant coupling was used to
motivate a general scaling
scenario for the quantum transition, conjectured to be valid even in the
region of runaway flows to
strong coupling.

We conclude by discussing implications for some related problems. It would
clearly be desirable
to extend our results to the case of insulating spin glasses of true
quantum Heisenberg spins. Unlike
the case for quantum rotors, the different vector components of a
Heisenberg spin do not commute with
each other---this leads to Berry phase terms in the path integral which
cannot be removed by any
obvious choice of variables, and complicates the problem substantially. A
Heisenberg spin model with
random infinite-range interactions was studied recently~\cite{Heisenberg}
and dramatically different
behavior was found even at this ``mean-field'' level---the quantum
paramagnetic phase was
gapless, unlike the fully gapped quantum paramagnet in infinite-range
quantum rotor
model~\cite{miller,letter}. A starting point for further analysis might be
to obtain a suitable
Landau action functional whose minimization reproduces the properties of
the infinite-range quantum
Heisenberg model.

A great deal of work~\cite{mott} has appeared recently on another quantum
transition in the large
dimensionality limit: the metal-insulator transition in Mott-Hubbard type
models. Like the spin-glass
problem, the order parameter for this transition is the long-time limit of
a correlation
function---at short times the correlation is non-zero on both sides of the
transition. The techniques
developed in this paper could perhaps be helpful in extending the infinite
dimensionality results to
the metal insulator transition in finite dimension systems. While this
paper was being written, we
learnt of the work of Kirkpatrick and Belitz~\cite{belitz} on the Landau
theory of a metal-insulator
transition in random electronic systems. While we do not understand the
details of their analysis,
there does appear to be at least a passing resemblance of their methods to
ours---like us, they find
it necessary to consider a linear term in the order parameter, to which
randomness couples most
effectively.

\acknowledgements

The research was supported by NSF Grants No. DMR-91-57484 (NR),
DMR-92-24290 (SS and JY),
and DMR-91-15491 (JY).
We are pleased to thank D.S. Fisher, D.A. Huse, and A.P. Young for
useful discussions.

\appendix

\section{Derivation of the Landau effective action}
\label{derivation}

In this appendix we outline an explicit derivation of the Landau effective
action
${\cal A}$ in (\ref{landau}) from the Hamiltonians ${\cal H}_R$ and ${\cal
H}_I$.
It is slightly more convenient to work with soft spins rather than the
fixed-length
quantum rotors or Ising spins (although the derivation below can be
extended to these
cases). We begin with the path-integral for these spins in the presence of
fixed
realization of the disorder
\begin{eqnarray}
Z =\int {\cal D} S_{i \mu}&& \exp \left( - \int d \tau \left\{ {\cal L}_0
(S_{i\mu})
-\sum_{<ij>} J_{ij} S_{i\mu} S_{j \mu}
\right\}\right) \nonumber \\
{\cal L}_0 =&& \sum_i \left[ \frac{1}{2g} \left( \partial_{\tau} S_{i \mu}
\right)^2 +
\frac{m^2}{2} S_{i \mu}^2 + \frac{\tilde{u}}{2} \left(S_{i\mu}^2 \right)^2
\right] .
\end{eqnarray}
This action may be interpreted as that of $M$-component harmonic
oscillators on the
sites $i$ of a lattice, with a non-linear self-coupling $u$ and a random
interaction
$J_{ij}$. We now introduce replicas and average over symmetric distribution
of the
$J_{ij}$. Neglecting all but the second cumulant of the $J_{ij}$, we obtain
the replicated,
translationally invariant result
\begin{equation}
\left[Z^n \right] = \int {\cal D} S_{i \mu}^a \exp \left( - \int d \tau
\sum_a {\cal L}_0
(S_{i\mu}^a ) -\frac{J^2}{2} \sum_{<ij>} \int d \tau_1 d \tau_2 \sum_{ab}
 S_{i\mu}^a (\tau_1 ) S_{j \mu}^a (\tau_1 ) S_{i\mu}^b (\tau_2 ) S_{j
\mu}^b (\tau_2 )
\right) .
\end{equation}
Now, as in classical spin glasses~\cite{book}, we decouple the quartic term by
a
Hubbard Stratonovich transformation
\begin{eqnarray}
\left[ Z^n \right] = && \int {\cal D} Q_{i\mu\nu}^{ab}
\exp \left( - \frac{J^2}{2} \int d\tau_1 d \tau_2 \sum_{i,j} \sum_{ab}
 Q_{i\mu\nu}^{ab} (\tau_1 , \tau_2 )
K^{-1}_{ij} Q_{j\mu\nu}^{ab} (\tau_1 , \tau_2 )
 \right) Z_S \left[ Q \right]
\nonumber \\
Z_S \left[ Q \right] =\int && {\cal D} S_{i \mu}^a \exp  \left( - \int d \tau
\sum_a {\cal L}_0
(S_{i\mu}^a ) - \int d \tau_1 d \tau_2
\sum_i \sum_{ab} Q_{i\mu\nu}^{ab} ( \tau_1 , \tau_2
) S_{i\mu}^{a} (\tau_1 ) S_{i\nu}^{b} ( \tau_2 ) \right)
\end{eqnarray}
where $K_{ij}$ is the connectivity matrix of the lattice.
It is now straightforward to expand $Z_S [Q]$ in powers of
$Q_{\mu\nu}^{ab} (x, \tau_1 , \tau_2 ) - C \delta^{ab}_{\mu\nu} \delta
(\tau_1 - \tau_2 )$
for a suitably chosen value of $C$, as explained in
Section~\ref{landausec}; the constant $C$ can
then be absorbed into ${\cal L}_0$ and becomes part of the quadratic
$S_{i\mu}^2$
term which can be shown to remain stable at small $\tilde{u}$.
It is easy to see that
this procedure will generate all terms consistent with the criteria
enumerated in
Section~\ref{landausec}. In particular, the term linear in $Q$ appears
immediately.
The rather unfamiliar looking time derivatives in this linear term
can be seen to follow from a gradient expansion of a term
like $\int d^d x d \tau_1 d \tau_2 Q^{aa}_{\mu\mu} ( x, \tau_1, \tau_2)
f(\tau_1 -
\tau_2 )$ where $f(\tau) $ is an even function of $\tau$ which falls rapidly to
zero within a $\tau$ of order $1/g$.
All the terms in ${\cal A}$ in (\ref{landau}) will be generated at higher
orders,
with the exception of the last $1/t^2$ term.
There are two routes to generating such a term: \\
({\em i\/}) Renormalize the functional
integral over $Q$ itself.  At order $\kappa^2$ one observes from the
Feynman diagram in
Fig~\ref{twist} that an  effective $1/t^2$ term is generated. Thus, even
though it is absent
in the bare $Q$ action, it will eventually appear. For the renormalization
group
analysis, it is therefore advantageous to include the $1/t^2$ term at the
starting
point. \\
({\em ii\/}) Introduce additional {\em on-site} sources of randomness in
${\cal L}_0$.
Randomness in the value of $m^2$ couples to the $[S_{i\mu}^{a} (\tau )]^2$,
an operator which was the same quantum numbers as $Q^{aa} (x, \tau, \tau)$.
Integrating over
the randomness will then generate the $1/t^2$ term. Randomness in $g$ or
$\tilde{u}$ has the same
effect, eventually.

\section{Griffiths singularities in the quantum paramagnet}
\label{griffiths}
Our mean-field result (\ref{chi}) for the local susceptibility in the
paramagnetic phase has a gap at
low frequencies. As already noted, this feature is an artifact of the
mean-field theory and we expect
random fluctuations in the $J_{ij}$ to create local environments which will
have excitations within this
`gap'. A careful formulation of this ``Griffiths'' effect has been
presented by Thill and Huse for the
$M=1$ case~\cite{thill}: they found a power law $\sim \omega^{\phi}$
absorption at low frequency,
where the exponent $\phi$ varies continuously with microscopic couplings
and need not be positive.
Here we will extend their argument to $M>1$ and find a very different
result: there is only a much weaker
essential singularity in the absorption spectrum for rotors with a
continuous internal symmetry.

We will be satisfied here by sketching the basic idea: it should not be too
difficult to formalize the
argument below along the lines of Ref~\cite{thill}.
The important contribution to the long-time limit of the average, local,
spin correlation function in the
paramagnetic phase comes from regions whose local environment has couplings
similar to those in the
spin-glass phase. Let us examine the contribution of a region, ${\cal
R}_L$, of linear dimension $L$, whose
coupling constants are those required to be well within the spin-glass or
ferromagnetic phase. Such a region
will occur with a probability $\sim \exp(- c_1 L^d )$ where the constant
$c_1$ depends upon the precise
criteria chosen. Apart from short-lived fluctuations, the spins in ${\cal
R}_L$ will follow each other and
behave like a single block spin evolving in imaginary time.
Neglecting the coupling to the environment, the fluctuations of this block
spin can be described
by a one-dimensional (corresponding to imaginary time direction),
classical, $M$-component spin chain
with a ferromagnetic coupling $K \sim L^d$. Now the properties of such a
spin chain are well known:
it has a finite correlation `time' $\xi_{\tau}$ which scales with large
$K$, for $M >1 $, as
\begin{equation}
\xi_{\tau} \sim K \sim L^{d} .
\end{equation}
In contrast, for $M=1$ we have $\log \xi_{\tau} \sim K \sim
L^{d}$~\cite{thill}. This much shorter
correlation  time for continuous spins ($M>1$) is responsible for the
difference from the Ising case.
So for a typical site, $i$, in ${\cal R}_L$ we will have
\begin{equation}
\left\langle \hat{n}_i ( \tau ) \cdot \hat{n}_i (0) \right\rangle
\sim \exp ( - c_2 L^{-d} |\tau |)
\end{equation}
for some constant $c_2$. The above argument is equivalent to the statement
that the region ${\cal R}_L$
behaves like a single quantum rotor with coupling $g \sim L^{-d}$.

We can now add up the contribution of all regions like ${\cal R}_L$ and
obtain a lower-bound on the
average, local, spin correlation function:
\begin{eqnarray}
\left[ \left\langle \hat{n}_i ( \tau ) \cdot \hat{n}_i (0) \right\rangle
\right]
&\geq& c_3 \int d L \exp(- c_1 L^{d} ) \exp ( - c_2 L^{-d} |\tau | )
\nonumber \\
&\sim & c_4 \exp\left(-2 \sqrt{c_1 c_2 |\tau| } \right)
\end{eqnarray}
where the integral over $L$ was performed by the saddle-point method.
By an inverse Laplace transform, it can be shown that this leads to a low
frequency contribution to
$\chi^{\prime\prime} ( \omega )$ of
$\sim \mbox{sgn} ( \omega ) \exp(- c_1 c_2 / |\omega| )$.
So the gap in (\ref{chi}) is filled in by a weak absorption which has an
essential singularity at
$\omega=0$.

\begin{figure}
\caption{Phase diagram of the action ${\cal A}$ (Eqn
(\protect\ref{landau})) as a
function of temperature $T$ and the Landau parameter $r$ which is a measure of
the strength of
quantum fluctuations (for $M=1$, $r$ is proportional to the transverse
field). There is no
significance to the position of the $y$-axis {\em i.e.} $r=0$ does not
correspond to zero quantum
fluctuations which for (\protect\ref{hamrot}) and (\protect\ref{hamising})
is $g=0$. The full line
is the only phase transition and dashed lines denote crossovers between
different regimes.
The position of the phase transition
is given by $r=r_c (T)$ where $r_c (T) - r_c (0) \sim T^{1/z\nu}$
(in mean field theory $z\nu=1/2$ and
the position is given by (\protect\ref{defrct})).
We now list the characteristics of the regimes, and the conditions which
bound them
(note $r_c \equiv r_c (0)$):
(I) $(r-r_c )^{z\nu} \gg T$, quantum paramagnet: thermal effects are secondary;
(II) $|r-r_c |^{z\nu} \ll T$, `quantum-critical': the critical ground state
at $r=r_c$ and
its thermal excitations determine the physics;
(III) $| r- r_c (T) |^{z\nu} \ll T$, classical: the behavior similar to that of
the classical, finite-temperature, spin glass; and
(IV) $(r_c - r)^{z\nu} \gg T$, quantum spin-glass: as in I, thermal effects
are secondary,
but the ground state now has long-range, spin-glass order.
}
\label{phasediag}
\end{figure}

\begin{figure}
\caption{({\em a\/}) The double-line representation for the propagator of
the $Q$ field;
each line represents one of its constituent $S$ fields, and carries its own
replica ($a,b$),
vector $O(M)$ ($\mu,\nu$) and frequency ($\omega_1 , \omega_2$) indices.
Momentum ($k$)
is however only carried by the composite double-line propagator. Also shown
is a twisted
partner which can replace the untwisted propagator in all the Feynman diagrams
in Figs~\protect\ref{loops} and~\protect\ref{kappa}.
({\em b\/}) The quantum mechanical
interactions $u/t$ and $v/t$ with replica,
$O(M)$ and frequency indices shown explicitly. Note that the single lines
do preserve their
replica and $O(M)$ labels, but can exchange frequency. All double-line
propagators carry the same momentum.
({\em c\/}) The cubic coupling ${\kappa/t}$. Each single
line now preserves replica,
$O(M)$ and frequency labels through the vertex. The double lines however do
exchange
momenta, with the total momentum being conserved.
({\em d\/}) The randomness-induced $1/t^2$ quadratic coupling. Again, each
single line preserves
replica,
$O(M)$ and frequency labels.
}
\label{vertices}
\end{figure}

\begin{figure}
\caption{Mean-field phase diagram for the $M=1$ Ising spin glass in a
longitudinal field $h$.
The surface shown is the analog of the de Almeida-Thouless~\protect\cite{AT}
line; the equation (\protect\ref{ATres}) determines the surface position
close to quantum
critical point $r=r_c$ and $T=0$.  Replica symmetry breaking occurs for
values of $h$ below
the surface. However the strength of the replica symmetry breaking vanishes
both as
$h$ approaches the surface, and as $T \rightarrow 0$; the strength of the
replica symmetry
breaking is given by (\protect\ref{rsbAT}) }
\label{ATsurface}
\end{figure}

\begin{figure}
\caption{Phase diagram for the $M > 1$ rotor spin glass in a longitudinal
field $h$
at zero temperature. The boundary $h_{GT}$ is the quantum analog of the
Gabay-Toulouse
line~\protect\cite{book} and is given by
(\protect\ref{GTres}) in mean-field theory. The spin glass order parameters
$q_{L}$ and $q_{T}$
refer to replica off-diagonal components of $Q_{11}$ and $Q_{\mu\mu}$, $\mu
> 1$ respectively.
(Replica off-diagonal components of $Q_{\mu\nu}$ with $\mu\neq\nu$ are zero
everywhere).
The field $h$ points along the $1$ direction and couples linearly to the
rotor co-ordinate
$\protect\hat{n}$. Replica symmetry breaking is expected to occur in the
mean-field theory at
non-zero $T$ for all $h < h_{GT}$.
}
\label{GTphase}
\end{figure}

\begin{figure}
\caption{Phase diagram for the $M=3$ quantum rotor spin glass at $T=0$ in
the presence of a
field $H$ which couples to the conserved total angular momentum in the
$1,2$ plane.
The position of the boundary of the paramagnetic phase is given by
(\protect\ref{conserveres})
in mean-field theory. The finite $H$ spin glass/paramagnet quantum
transition is in a different
universality class from the $H=0$ transition. As in
Fig~\protect\ref{GTphase}, the spin-glass order
parameter $q$ denotes replica off-diagonal components of $Q$, and replica
off-diagonal components of
$Q_{\mu\nu}$ with $\mu\neq\nu$ are zero everywhere.
The boundary $H_3$ is given by (\protect\ref{H3res}) in mean-field theory.
Replica symmetry breaking is expected to
occur in the mean-field theory at non-zero $T$ everywhere in the spin glass
phase.}
\label{conservefig}
\end{figure}

\begin{figure}
\caption{ Diagrams contributing to the one-loop renormalization group
equations.
The diagrams contribute to the renormalization of
({\em a\/}) $r/\kappa t$, ({\em b\/}) $u/t$, $v/t$, $v/t$, $1/t^2$, and
$1/t^2$ respectively and
({\em c\/}) $\eta$, $1/t^2$, $v/t$, and $u/t$ respectively. Not shown are
the diagrams with
double-line propagators replaced by their twisted partners (See
Fig~\protect\ref{vertices}a). }
\label{loops}
\end{figure}
\begin{figure}
\caption{As in Fig~\protect\ref{loops}; diagram contributing to the
renormalization of
$\kappa/t$. }
\label{kappa}
\end{figure}

\begin{figure}
\caption{Renormalization group flows in the $U,\kappa$ plane (recall
$U\equiv u + Mv$)
for $d > 8$ (we chose $\epsilon = 8-d=-1$).
The filled circles represent fixed points, and are labeled in the notation of
Eqn~(\protect\ref{rgfixedpoints}). The fixed points $B,C$ coincide in this
plane, but have
different fixed points values of $u$ (similarly for $D,E$). The only stable
fixed point is
$A$ and its basin of attraction is restricted to a portion of the $U>0$
quadrant. The remainder of the quadrant, and all $U<0$, eventually have
runaway flows to
strong coupling. The flow into the stable fixed point $A$ remains marginal
in the $U$
direction for {\em all\/} $d > 8$.
}
\label{rgflowdgt8}
\end{figure}

\begin{figure}
\caption{
As in Fig~\protect\ref{rgflowdgt8} but for $d=8$ ($\epsilon=0$). The only
fixed point is
$\kappa=U=0$, and its domain of attraction is $\kappa^2 \protect\leq U/20$.
For other initial
values we have runaway flow to strong coupling. All the fixed points in
Fig~\protect\ref{rgflowdgt8} coalesce into the $y=\kappa=0$ fixed point as
$d$ approaches 8
from above.
}
\label{rgflowdeq8}
\end{figure}

\begin{figure}
\caption{As in Fig~\protect\ref{rgflowdgt8} but for $d<8$ (we chose
$\epsilon=8-d=0.2$).
Only the
fixed point $A$ is now physical as the others have imaginary value of
$\kappa$. However,
$A$ is unstable for any non-zero $\kappa$, and all physical couplings flow to
strong coupling. }
\label{rgflowdlt8}
\end{figure}

\begin{figure}
\caption{Feynman diagram which generates an effective coupling $1/t^2$ even if
only the cubic coupling $\kappa/t$ is present originally.}
\label{twist}
\end{figure}


\begin{references}

\bibitem{kogut} J.B. Kogut, Rev. Mod. Phys. {\bf 51}, 659 (1979).

\bibitem{chn} S. Chakravarty, B.I. Halperin, and D.R. Nelson, Phys. Rev. B
{\bf 39}, 2344 (1989).

\bibitem{csy} A. Chubukov, S. Sachdev, and J. Ye, Phys. Rev. B {\bf 49},
11919 (1994).

\bibitem{rosen} W. Wu, B. Ellman, T.F. Rosenbaum, G. Aeppli, and D.H. Reich,
Phys. Rev. Lett. {\bf 67}, 2076 (1991); W. Wu, D. Bitko, T.F. Rosenbaum,
and G. Aeppli, Phys. Rev. Lett. {\bf 71}, 1919 (1993).

\bibitem{qcrit} S. Sachdev and J. Ye, Phys. Rev. Lett. {\bf 69}, 2411 (1992).

\bibitem{haldane} F.D.M. Haldane, Phys. Lett. {\bf 93A}, 464 (1983); Phys. Rev.
Lett. {\bf 50}, 1153 (1983); J. Appl. Phys. {\bf 57}, 3359 (1985).

\bibitem{youngbinder} K. Binder and A.P. Young, Rev. Mod. Phys. {\bf 58},
801 (1986).

\bibitem{book} {\em Spin Glasses} by K.H. Fischer and J.A. Hertz, Cambridge
University Press, Cambridge (1991).

\bibitem{wu} B.M. McCoy and T.T. Wu, Phys. Rev.
{\bf 176}, 631 (1968); B.M. McCoy, Phys. Rev. B {\bf 2}, 2795 (1970).

\bibitem{shankar} R. Shankar and G. Murthy, Phys. Rev. B {\bf 36}, 536 (1987).

\bibitem{daniel} D.S. Fisher, Phys. Rev. Lett. {\bf 69}, 534 (1992);
Harvard University preprint.

\bibitem{doro} S.N. Dorogovstev, Phys. Lett. {\bf 76A}, 169 (1980).

\bibitem{cardy} D. Boyanovsky and J.L. Cardy, Phys. Rev. B {\bf 26}, 154
(1982).

\bibitem{lawrie} D. Lawrie and V.V. Prudnikov, J. Phys. C {\bf 17}, 1655
(1984).

\bibitem{BrayMoore} A.J. Bray and M.A. Moore, J. Phys. C {\bf 13}, L655 (1980).

\bibitem{Gold} Y.Y. Goldschmidt and P.-Y. Lai, Phys. Rev. Lett. {\bf 64},
2467 (1990);
K.D. Usadel, G. Buttner, and T.K. Kopec, Phys. Rev. B {\bf 44}, 12583 (1990);
B. Boechat, R.R. dos Santos, and M.A. Continentino, Phys. Rev. B {\bf 49},
6404 (1994);
T.K. Kopec Phys. Rev. B {\bf 50}, 9963 (1994).

\bibitem{miller} D.A. Huse and J. Miller, Phys. Rev. Lett. {\bf 70}, 3147
(1993).

\bibitem{letter} J. Ye, S. Sachdev, and N. Read, Phys. Rev. Lett.
{\bf 70}, 4011 (1993); J. Ye, Ph.D. thesis, Yale University, unpublished.

\bibitem{guo} M. Guo, R.N. Bhatt, and D.A. Huse, Phys. Rev. Lett. {\bf 72},
4137 (1994).

\bibitem{rieger} H. Rieger and A.P. Young, Phys. Rev. Lett. {\bf 72}, 4141
(1994).

\bibitem{rieger2} A. Crisanti and H. Rieger, preprint, cond-mat/9406006.

\bibitem{opperman} R. Oppermann and A. Muller-Groeling, Nucl. Phys. B {\bf
401}, 507 (1993);
R. Oppermann and M. Binderberger, Ann. Physik {\bf 3}, 494 (1994).

\bibitem{thill} M.J. Thill and D.A. Huse, preprint.

\bibitem{private_huse} D. Huse has pointed out that the behavior of
$\chi_{nl}$ in the limit of
infinite $d$ is different from that in the infinite-range model; our mean
field theory presumably
describes the infinite $d$ limit.

\bibitem{ludwig} B. Duplantier and A.W.W. Ludwig, Phys. Rev. Lett. {\bf 66},
247 (1991); A.W.W. Ludwig, Nucl. Phys. {\bf B330}, 639 (1990).

\bibitem{yang} C.N. Yang and T.D. Lee, Phys. Rev. {\bf 87}, 404 (1952);
T.D. Lee and C.N. Yang, Phys. Rev. {\bf 87}, 410 (1951).

\bibitem{mef} M.E. Fisher, Phys. Rev. Lett. {\bf 40}, 1610 (1978).

\bibitem{jlc} J.L. Cardy, Phys. Rev. Lett. {\bf 54}, 1354 (1985).

\bibitem{thouless} J.R.L. de Almeida, R.C. Jones, J.M. Kosterlitz, and D.J.
Thouless,
J. Phys. C {\bf 11}, L871 (1978).

\bibitem{pat} G. Parisi and G. Toulouse, J. Phys. (Paris) Lett. {\bf 41},
L361 (1980)

\bibitem{temesvari} T. Temesv\'{a}ri, J. Phys. A {\bf 22}, L1025 (1989).

\bibitem{AT} J.R.L. de Almeida and D.J. Thouless,
J. Phys. A {\bf 11}, 983 (1978).

\bibitem{mezard} M. Mezard, G. Parisi, N. Sourlas, G. Toulouse, and M.
Virasoro,
J. Phys. (Paris) {\bf 45}, 843 (1984); M. Mezard, G. Parisi and M.
Virasoro, J. Phys. (Paris) Lett.
{\bf 46}, L217 (1985).

\bibitem{conserve} S. Sachdev, Z. Phys. B {\bf 94}, 469 (1994).

\bibitem{fishersigma} D.S. Fisher, Phys. Rev. B {\bf 39}, 11783 (1989).

\bibitem{geoff} G. Grinstein, Phys. Rev. Lett. {\bf 37}, 944 (1976).

\bibitem{dsf} D.S. Fisher, Phys. Rev. Lett. {\bf 56}, 416 (1986).

\bibitem{schwartz} M. Schwartz and A. Soffer, Phys. Rev. Lett. {\bf 55},
2499 (1985).

\bibitem{schwartz2} M. Schwartz, M. Gofman, and T. Natterman, Physica A
{\bf 178}, 6 (1991);
M. Gofman, J. Adler, A. Aharony, A.B. Harris and M. Schwartz, Phys. Rev. Lett.
{\bf 71}, 1569 (1993) and references therein.

\bibitem{harris} A.B. Harris, J. Phys. C {\bf 7}, 1671 (1974);
J.T. Chayes, L. Chayes, D.S. Fisher, and T. Spencer, Phys. Rev. Lett.
{\bf 57}, 2999 (1986).

\bibitem{Heisenberg} S. Sachdev and J. Ye, Phys. Rev. Lett. {\bf 70}, 3339
(1993).

\bibitem{mott} G. Kotliar in {\em Strongly Correlated
Electronic Materials\/} edited by
 K. S. Bedell, Z. Wang, D. Meltzer, A. Balatsky, and
E Abrahams, Addison Wesley (1993) and references therein.


\bibitem{belitz} T.R. Kirkpatrick and D. Belitz, cond-mat/9408009.

\end{references}
\end{document}